\PassOptionsToPackage{table}{xcolor}
\PassOptionsToPackage{bookmarksnumbered,unicode,ocgcolorlinks,pdftex}{hyperref}
\documentclass[acmsmall,10pt,screen]{acmart}
\usepackage{etex}
\usepackage{suffix}
\usepackage{okgeneralmath}
\usepackage{okcategory}
\usepackage{bbm}
\usepackage{multicol}
\usepackage{mathtools}
\usepackage{mathpartir}
\usepackage{oksemantics}
\usepackage{commath}
\usepackage{listings}
\usepackage[inline]{enumitem}
\usepackage{wrapfig}
\renewcommand\mathbf\textbf

\newcommand{\hy}[1]{
\smallskip\noindent\fbox{\begin{minipage}{.97\linewidth}{\bf HY:}
{\rm #1}\end{minipage}}}

\newcommand{\ok}[1]{
\smallskip\noindent\fbox{\begin{minipage}{.97\linewidth}{\bf OK:}
{\rm #1}\end{minipage}}}

\newcommand{\commentout}[1]{}

\newcommand\seclabel[1]{\label{sec:#1}}
\renewcommand\secref[1]{Sec.~\ref{sec:#1}}
\newcommand\subseclabel[1]{\label{subsec:#1}}
\newcommand\subsecref[1]{\S~\ref{subsec:#1}}
\newcommand\subsubseclabel[1]{\label{subsubsec:#1}}
\newcommand\subsubsecref[1]{\S~\ref{subsubsec:#1}}
\newcommand\figlabel[1]{\label{fig:#1}}
\renewcommand\figref[1]{Fig.~\ref{fig:#1}}

\newcommand\AddReferences[2]{
  \expandafter\newcommand\csname#1ref\endcsname[1]{#2~\ref{#1:##1}}
  \expandafter\newcommand\csname#1label\endcsname[1]{\label{#1:##1}}
  \WithSuffix\expandafter\newcommand\csname#1ref\endcsname*[1]{\ref{#1:##1}}
  \WithSuffix\expandafter\newcommand\csname#1label\endcsname+[1]{\hypertarget{#1+:##1}{}\zref@labelbyprops{#1:##1}{oktheoremfreetext}}
  \WithSuffix\expandafter\newcommand\csname#1ref\endcsname+[1]{\hyperlink{#1+:##1}{{{\let\ref\@refstar#2~\zref@extract{#1:##1}{oktheoremfreetext}}}}}
  \WithSuffix\expandafter\newcommand\csname#1ref\endcsname-[1]{\hyperlink{#1+:##1}{{\let\ref\@refstar{\zref@extract{#1:##1}{oktheoremfreetext}}}}}
}

\AddReferences{example}{Ex.}
\AddReferences{lemma}{Lemma}
\AddReferences{theorem}{Thm.}
\AddReferences{definition}{Def.}
\newcommand\defref\definitionref
\newcommand\deflabel\definitionlabel

\newcommand{\citepos}[1]{\citeauthor{#1}'s~\citeyearpar{#1}}
\WithSuffix\newcommand\citepos*[2]{\citeauthor{#1}'s #2~\citeyearpar{#1}}
\WithSuffix\newcommand\citepos-[2]{\citeauthor{#1}#2~\citeyearpar{#1}}

\newcounter{countitems}
\newcounter{nextitemizecount}
\newcommand{\setupcountitems}{
  \stepcounter{nextitemizecount}
  \setcounter{countitems}{0}
  \preto\item{\stepcounter{countitems}}
}
\makeatletter
\newcommand{\computecountitems}{
  \edef\@currentlabel{\number\c@countitems}
  \label{countitems@\number\numexpr\value{nextitemizecount}-1\relax}
}
\newcommand{\nextitemizecount}{
  \getrefnumber{countitems@\number\c@nextitemizecount}
}
\newcommand{\previtemizecount}{
  \getrefnumber{countitems@\number\numexpr\value{nextitemizecount}-1\relax}
}
\makeatother

\definecolor{shade}{RGB}{223,223,223}
\definecolor{unshade}{RGB}{255,255,255}

\usepackage{tcolorbox}
\newtcbox{\shadebox}{on line,arc=1pt, outer arc=2pt,
  colback=shade,colframe=shade,boxsep=0pt,
  left=1pt,right=1pt,top=2pt,bottom=2pt,
  boxrule=0pt,bottomrule=1pt,toprule=1pt}
\newcommand{\shade}[1]{
        \shadebox{\ensuremath{#1}}
}
\newtcbox{\unshadebox}{on line,arc=1pt, outer arc=2pt,
  colback=unshade,colframe=shade,boxsep=0pt,
  left=1pt,right=1pt,top=2pt,bottom=2pt,
  boxrule=0pt,bottomrule=1pt,toprule=1pt}

\newcommand\syncat[1]{\mspace{-25mu}\synname{#1}}
\newcommand\synname[1]{\qquad\text{#1}}

\newenvironment{syntax}[1][]{
\(
  \rowcolors{100}{white}{white}
  \begin{array}[t]{#1l@{\quad\!\!}*3{l@{}}@{\,}l}
}{
\end{array}
\)
}

\newenvironment{sugar}{
  \(
  \begin{array}[t]{@{\bullet\ }l@{\mspace{25mu}}ll}
    \multicolumn{1}{@{\hphantom{\bullet}}l}{\text{Sugar}} & \text{Elaboration}\\
    \hline
}{
  \end{array}
  \)
}

\newcommand\gdefinedby{::=}
\newcommand\gor{\mathrel{\lvert}}
\makeatletter

\newcommand\@TyAlph[1]{
\ifcase #1\or \tau\or \sigma\or \rho\else \@ctrerr \fi
}
\newcommand\ty[1][1]{{\@TyAlph{#1}}}

\newcommand\tvar[1][1]{{\@TyVarAlph{#1}}}
\newcommand\@TyVarAlph[1]{
\ifcase #1\or \alpha\or \beta\or \gamma\else \@ctrerr \fi
}

\newcommand\Functor[1][1]{\TypeCons{\@FunAlph{#1}}}
\newcommand\@FunAlph[1]{
\ifcase #1\or F\or G\or H\else \@ctrerr \fi
}

\newcommand\Type[1][1]{{\mathrm{\@TypeAlph{#1}}}}
\newcommand\@TypeAlph[1]{
\ifcase #1\or A\or B\or C\else \@ctrerr \fi
}

\newcommand\var[1][1]{{\@VarAlph{#1}}}
\newcommand\@VarAlph[1]{
\ifcase #1\or x\or y\or z\or u\or v\or w\else \@ctrerr \fi
}

\newcommand\trm[1][1]{{\@TermAlph{#1}}}
\newcommand\@TermAlph[1]{
\ifcase #1\or t\or s\or r\else \@ctrerr \fi
}

\newcommand\ir[1][1]{\monad{\uir[#1]}}

\newcommand\uir[1][1]{\mathop{{}\@IRAlph{#1}}\nolimits}
\newcommand\@IRAlph[1]{
\ifcase #1\or T\or S\or R\else \@ctrerr \fi
}

\newcommand\IF[1][1]{\monad{\uIF[#1]}}
\newcommand\uIF[1][1]{\mathop{{}\@IFAlph{#1}}\nolimits}
\newcommand\@IFAlph[1]{
\ifcase #1\or F\or G\or H\else \@ctrerr \fi
}

\newcommand\tfm[1][1]{\mathop{{}\underline{\trm[#1]}}\nolimits}

\newcommand\cat[1][1]{\Cat{\@CatAlph#1}}
\newcommand\@CatAlph[1]{
\ifcase #1\or C\or D\or A\or B\or E\else \@ctrerr \fi
}

\newcommand\F[1][1]{\@FunctorAlph#1}
\newcommand\@FunctorAlph[1]{
\ifcase #1\or F\or G\or H\\else \@ctrerr \fi
}

\newcommand\nt[1][1]{\@NatTransAlph#1}
\newcommand\@NatTransAlph[1]{
\ifcase #1\or \alpha\or \beta\or \gamma\else \@ctrerr \fi
}

\newcommand\alg[1][1]{\@AlgAlph#1}
\newcommand\@AlgAlph[1]{
\ifcase #1\or A\or B\or C\else \@ctrerr \fi
}

\newcommand\qmu[1][1]{\underline{\@MuAlph#1}}
\newcommand\@MuAlph[1]{
\ifcase #1\or \mu\or \nu\or \xi\or \zeta\else \@ctrerr \fi
}

\makeatother

\newcommand\Cns{\ell}
\newcommand\Inj[2][\,]{\mathrm{#2}#1}
\newcommand\Variant[1]{\{ #1 \}}
\newcommand\vor{\mathrel{\big\lvert}}
\newcommand\Unit{\textbf{1}}
\WithSuffix\newcommand\t*{*}
\newcommand\rec[2]{\mu#1.#2}
\newcommand\To{\to}
\newcommand\ctx{\Gamma}
\newcommand\Tyctx{\Delta}
\newcommand\kinf{\vdash_{\mathrm k}}
\newcommand\Dkinf[2][]{\Tyctx#1\kinf#2 : \typ}
\newcommand\typ{\mathrm{type}}
\newcommand\context{\mathrm{context}}

\newcommand\TypeCons[1]{\mathop{{}\mathrm{#1}}\nolimits}

\newcommand\List{\TypeCons{List}}
\newcommand\boolty{{\TypeCons{bool}}}
\newcommand\tru{\mathop{\Inj[]{True}}}
\newcommand\fls{\mathop{\Inj[]{False}}}

\newcommand\I{\mathbb{I}}
\newcommand\ereals{\overline\reals}
\WithSuffix\newcommand\ereals+{\overline\reals_+}
\WithSuffix\newcommand\reals+{\reals_+}

\newcommand\tInj[3][\,]{#2.#3#1}
\newcommand\tUnit{()}
\newcommand\tpair[2]{(#1, #2)}

\newcommand\troll[2][\,]{#2.\textbf{roll}#1}
\newcommand\fun[1]{\lambda #1.\,}
\newcommand\pfun[1]{
  \lambda\,\begin{array}[t]{@{}l@{}l@{}l@{}}
           \!\{\,
           #1
           \}
         \end{array}}
\newcommand\vmatch[2]{\textbf{match}\,#1\,\textbf{with}\,\{#2\}}
\newcommand\pmatch[4]{\textbf{match}\,#1\,\textbf{with}\,\tpair{#2}{#3}\To#4}
\newcommand\rmatch[4][\,]{\textbf{match}\,#2\,\textbf{with}#1\roll#3\To#4}
\newcommand\imatch[2]{\begin{array}[t]{@{}l@{\,}l}
    \vmatch{&#1}{\\
    & \begin{array}[t]{@{}l@{\To}l}
        #2
      }\end{array}
    \end{array}}
\newcommand\tfold[2][\,]{#2.\textbf{fold}#1}

\newcommand\tcts{\phi}
\newcommand\roll[1][\,]{\textbf{roll}#1}

\newcommand\fold{\textbf{fold}\,}

\newcommand\expfun[2]{\lambda #1:#2.}
\newcommand\letin[3]{\textbf{let}\,#1=#2\,\textbf{in}\,#3}
\newcommand\expletin[2]{\letin{(#1:#2)}}
\newcommand\tif[3]{\mathrel{\textbf{if}}#1\mathrel{\textbf{then}}#2\mathrel{\textbf{else}}#3}

\newcommand\foldlist[2]{\mathrm{foldr}\,#1\,#2}
\WithSuffix\newcommand\++{\mathbin{+\kern-.4em+}}
\newcommand\tmap[1][\,]{\mathrm{map}#1}

\newcommand\lst[1]{[#1]}
\WithSuffix\newcommand\lst-{\lst{\ }}
\newcommand\aggr{\mathop{\mathrm{aggr}}\nolimits}
\newcommand\add{\mathop{\mathrm{add}}\nolimits}
\newcommand\waggr{\mathop{\mathrm{waggr}}\nolimits}
\newcommand\sequence{\mathop{\mathrm{sequence}}\nolimits}
\newcommand\concat{\mathop{\mathrm{concat}}\nolimits}
\newcommand\replicate{\mathop{\mathrm{replicate}}\nolimits}

\newcommand\tinf{\vdash}
\newcommand\Ginf[3][]{\ctx #1\tinf #2 : #3}
\newcommand\subst[2]{#1{}[#2]}
\newcommand\sfor[2]{#1 \mapsto #2}

\newcommand\scomment[1]{\texttt{--}\ \text{#1}}
\newcommand\gobble[1]{}

\newcommand\sem\scottbrackets
\newcommand\carrier[1]{\left\lvert#1\right\rvert}
\renewcommand\terminal{\mathbbold{1}}
\newcommand\Two{\mathbbold{2}}
\DeclareMathAlphabet{\mathbbold}{U}{bbold}{m}{n}
\renewcommand\mathbbold\underline
\renewcommand\initial{\mathbbold{0}}
\newcommand\zero{\mathbbold{Z}}
\newcommand\zeromor{\textbf{0}}

\newcommand\algfold[1][]{\mathrm{fold}^{#1}}
\newcommand\codiag{\nabla}

\newcommand\Base{\mathcal{B}\sem}
\newcommand\tenv{d}

\usepackage{extpfeil}
\newextarrow{\xtoto}{{20}{20}{20}{20}}
   {\bigRelbar\bigRelbar{\bigtwoarrowsleft\rightarrow\rightarrow}}
\makeatletter
\newbox\xrat@below
\newbox\xrat@above
\newcommand{\xrightarrowtail}[2][]{
  \setbox\xrat@below=\hbox{\ensuremath{\scriptstyle #1}}
  \setbox\xrat@above=\hbox{\ensuremath{\scriptstyle #2}}
  \pgfmathsetlengthmacro{\xrat@len}{max(\wd\xrat@below,\wd\xrat@above)+.6em}
  \mathrel{\tikz [>->,baseline=-.75ex]
                 \draw (0,0) -- node[below=-2pt] {\box\xrat@below}
                                node[above=-2pt] {\box\xrat@above}
                       (\xrat@len,0) ;}}
\makeatletter
\usepackage{tikz}
\newcommand{\xdasharrow}[2][->]{
\tikz[baseline=-\the\dimexpr\fontdimen22\textfont2\relax]{
\node[anchor=south,font=\scriptsize, inner ysep=1.5pt,outer xsep=2.2pt](x){$ #2$};
\draw[shorten <=3.4pt,shorten >=3.4pt,dashed,#1](x.south west)--(x.south east);
}
}
\makeatother
\newcommand\monad\underline
\newcommand\return{\mathop{\mathrm{return}}\nolimits}
\newcommand\sreturn{\mathop{\textbf{return}}\nolimits}
\newcommand{\bind}{\mathrel{\scalebox{0.8}[1]{\(>\!\!>\!=\)}}}
\newcommand\GenMonad[4]{
    \begin{array}[t]{@{}l@{}l@{}l@{}l@{}}
      \textbf{in}&\multicolumn{2}{@{}l@{}}{\textbf{stance}\hphantom\ \mathrm{#1}\, (#3)\, \textbf{where}\qquad} #2\\
                 #4
    \end{array}
  }
\newcommand\GenMonadBreak[4]{
    \begin{array}[t]{@{}l@{}l@{}l@{}l@{\qquad}l@{}l@{}l@{}l@{}l@{}l@{}l@{}l}
      \textbf{in}&\multicolumn{2}{@{}l@{}}{\textbf{stance}\hphantom\ \mathrm{#1}\, (#3)\, \textbf{where}\qquad} #2\\[3pt]
                 #4
    \end{array}
  }
\newcommand\Monad[4][]{
  \GenMonad{Monad}{#1}{#2}{&#3\\&#4}
}
\newcommand\DiscMonad[7][]{
  \GenMonad{Discrete\ Monad}{#1}{#2}
    {&#3\\&#4\\&#5\\&#6\\&#7}
}

\newcommand\SampleMonad[6][]{
  \GenMonad{Sampling\ Monad}{#1}{#2}
    {&#3\\&#4\\&#5\\&#6}
}

\newcommand\DiscTransBreak[9][]{
  \GenMonadBreak{Inf\ Trans}{#1}{#2}
           {&#3&&\\&#4&\\&#5\\&#6\\&#7\\&#8\\&#9}
}

\newcommand\RepTrans[9][]{
  \GenMonad{\FromKind{#2}#3\ Trans}
           {#1}
           {#4}
           {&#5\\&#6\\&#7\\&#8\\&#9}
}

\newcommand\FromKind[1]{
  \ifthenelse{\equal{#1}{Rep}}
             {}
             {#1\implies}
}

\WithSuffix\newcommand\<-{\mathrel{\leftarrow}}
\newcommand\domon[1]{\textbf{do}\,\{#1\}}
\newcommand\expdomon[2]{#1.\domon{#2}}

\newcommand\sflip{\textbf{flip}}
\newcommand\sscore{\mathop{\textbf{score}}\nolimits}

\newcommand\flip{\mathrm{flip}}
\newcommand\score{\mathop{\mathrm{score}}\nolimits}

\newcommand\Mass{\mathop{\mathrm{Mass}}}

\newcommand{\support}{\mathop{\mathrm{supp}}\nolimits}
\newcommand\constantly\underline

\newcommand\mean{\mathop{{}m}\nolimits}

\newcommand\Mea{\mathop{\mathrm{{}M}}\nolimits}
\newcommand\Prob{\mathop{\mathrm{{}P}}\nolimits}
\newcommand\sub{\mathop{\mathrm{sub}}\nolimits}

\newcommand\Qmonad{\Mea}

\newcommand\Term{\mathop{\mathrm{Term}}\nolimits}
\newcommand\Enum{\mathop{\mathrm{Enum}}\nolimits}
\newcommand\DSam{\mathop{\mathrm{DSam}}\nolimits}
\newcommand\Sam{\mathop{\mathrm{Sam}}\nolimits}

\newcommand\half{\tfrac12}

\newcommand\Writer{\mathop{{}\mathrm{W}}\nolimits}
\newcommand\itmap{\mathop{\mathrm{tmap}}\nolimits}
\newcommand\stmap{\mathop{\textbf{tmap}}\nolimits}
\newcommand\lift{\mathop{\mathrm{lift}}\nolimits}
\newcommand\slift{\mathop{\textbf{lift}}\nolimits}

\newcommand\ListT{\mathop{{}\mathrm{ListT}}\nolimits}
\newcommand\Pop{\mathop{{}\mathrm{Pop}}\nolimits}
\newcommand\Sus{\mathop{{}\mathrm{Sus}}\nolimits}
\newcommand\Seq\Sus
\newcommand\WSam{\mathop{{}\mathrm{WSam}}\nolimits}

\newcommand\noassumptions[1]{
  \multicolumn{1}{@{}l@{}}{}&\multicolumn{1}{@{}l@{\definedby{}}}{
    #1
  }
}
\newenvironment{notation}
{
  \[
    \begin{array}{@{}l@{{}\tinf{}}l@{{}\definedby{}}l@{\ \ }l}
      \multicolumn{1}{@{}l@{}}{}
      &
      \multicolumn{1}{@{}l@{\ }}{
      \text{Notation}
      }
      &
      \multicolumn{1}{@{}l}{
        \text{Meaning}
      }
      &
      \multicolumn{1}{@{}l}{
        \text{Terminology}
      }
      \\
      \hline
}{
  \end{array}
  \]
  \ignorespacesafterend
}
\newcommand\qdelta{\underline\delta}
\DeclareSymbolFont{largesymbolsA}{U}{txexa}{m}{n}
\DeclareMathSymbol{\sqintop}{\mathop}{largesymbolsA}{14}
\DeclareMathSymbol{\sqiintop}{\mathop}{largesymbolsA}{80}
\DeclareMathSymbol{\sqiiintop}{\mathop}{largesymbolsA}{82}

\newcommand\qint{\sqintop}
\newcommand\qiint{\sqiintop}

\newcommand\Exp[1][]{\mathop{\mathbb{E}^{\mathrlap{#1}}}}
\newcommand\sampled{\sim}

\newcommand\absc{\mathrel{<\kern-.4em<}}
\newcommand\der[2]{\tfrac{\dif #1}{\dif #2}}

\newcommand{\braid}[1][\ ]{\mathrm{swap}#1}
\newcommand\bprod{\boxtimes}

\newcommand\Sampling[1][\ ]{\mathrm{Sam}#1}

\newcommand\Qbs{\Category{Qbs}}
\newcommand\sample{\textbf{sample}}
\newcommand\ssample{\mathrm{sample}}
\newcommand\Uniform[1][]{\textbf{U}_{#1}}
\newcommand\UniformD{\mathrm{Uniform}}

\newcommand\spawn{\mathop{\mathrm{spawn}}\nolimits}
\newcommand\dwrand{\mathop{\mathrm{dwrand}}\nolimits}
\newcommand\dwsampler{\mathop{\mathrm{dwsample}}\nolimits}
\newcommand\resample{\mathop{\mathrm{resample}}\nolimits}
\newcommand\fuel{\textit{fuel}}

\newcommand\result{\textit{result}}

\newcommand\fail{\mathrm{fail}}

\newcommand\spark{\mathop{\mathrm{spark}}\nolimits}
\newcommand\finish{\mathop{\mathrm{finish}}\nolimits}
\newcommand\adva{\mathop{\mathrm{advance}}\nolimits}

\newcommand\smc{\mathop{\mathrm{smc}}\nolimits}
\newcommand\rmsmc{\mathop{\mathrm{rmsmc}}\nolimits}

\newcommand\mh[1][\pert]{\eta_{#1}}
\newcommand\paths{\mathop{\mathrm{Paths}}\nolimits}
\newcommand\weight[1]{w_{#1}}
\newcommand\valuate[1]{v_{#1}}
\newcommand\Trace{\mathop{\mathrm{Tr}}\nolimits}
\newcommand\iprior{\mathop{\mathrm{pri}}\nolimits}
\newcommand{\UD}{\mathrm{U_D}}
\newcommand\marginal{\mathop{\mathrm{marginal}}\nolimits}

\newcommand\isubterm{\mathop{\mathrm{sub}}\nolimits}

\newcommand\itake{\mathop{\mathrm{take}}\nolimits}

\usepackage{booktabs}
\usepackage{subcaption}

\setcopyright{rightsretained}
\acmPrice{}
\acmDOI{10.1145/3158148}
\acmYear{2018}
\copyrightyear{2018}
\acmJournal{PACMPL}
\acmVolume{2}
\acmNumber{POPL}
\acmArticle{60}
\acmMonth{1}

\begin{document}

\title{Denotational Validation of Higher-Order Bayesian Inference}
 \author{Adam \'Scibior}
\email{ams240@cam.ac.uk}
\affiliation{
  \department{Department of Engineering}
  \institution{University of Cambridge}
  \streetaddress{Trumpington Street}
  \city{Cambridge}
  \postcode{CB2 1PZ}
  \country{England}
}
\affiliation{
  \department{Empirical Inference Deparment}
  \institution{Max Planck Institute for Intelligent Systems}
  \streetaddress{Spemannstrasse 34}
  \city{T\"ubingen}
  \postcode{72076}
  \country{Germany}
}
\author{Ohad Kammar}
\orcid{0000-0002-2071-0929}
\email{ohad.kammar@cs.ox.ac.uk}
\affiliation{
  \institution{University of Oxford}
  \department{Department of Computer Science}
  \streetaddress{Wolfson Building, Parks Road}
  \city{Oxford}
  \postcode{OX1 3QD}
  \country{England}
}
\author{Matthijs V\'ak\'ar}
\email{matthijs.vakar@magd.ox.ac.uk}
\affiliation{
  \institution{University of Oxford}
  \department{Department of Computer Science}
  \streetaddress{Wolfson Building, Parks Road}
  \city{Oxford}
  \postcode{OX1 3QD}
  \country{England}
}
\author{Sam Staton}
\email{sam.staton@cs.ox.ac.uk}
\affiliation{
  \institution{University of Oxford}
  \department{Department of Computer Science}
  \streetaddress{Wolfson Building, Parks Road}
  \city{Oxford}
  \postcode{OX1 3QD}
  \country{England}
}
\author{Hongseok Yang}
\email{Hongseok.Yang@cs.ox.ac.uk}
\affiliation{
  \institution{KAIST}
  \country{South Korea}
}
\author{Yufei Cai}
\email{yufei.cai@uni-tuebingen.de}
\affiliation{
  \institution{Universit\"at T\"ubingen}
  \department{Programming Languages and Software Technology}
  \streetaddress{Sand 13}
  \city{T\"ubingen}
  \postcode{72076}
  \country{Germany}
}
\author{Klaus Ostermann}
\email{klaus.ostermann@uni-tuebingen.de}
\affiliation{
  \institution{Universit\"at T\"ubingen}
  \department{Programming Languages and Software Technology}
  \streetaddress{Sand 13}
  \city{T\"ubingen}
  \postcode{72076}
  \country{Germany}
 }
\author{Sean K.~Moss}
\email{S.K.Moss@dpmms.cam.ac.uk}
\affiliation{
   \department{Department of Pure Mathematics and Mathematical Statistics}
   \institution{University of Cambridge}
   \streetaddress{Centre for Mathematical Sciences, Wilberforce Road}
   \city{Cambridge}
   \postcode{CB3 0WB}
   \country{England}
}
\affiliation{
   \department{University College}
   \institution{University of Oxford}
   \streetaddress{High Street}
   \city{Oxford}
   \postcode{OX1 4BH}
   \country{England}
}
\author{Chris Heunen}
\email{chris.heunen@ed.ac.uk}
\affiliation{
  \department{School of Informatics}
  \institution{University of Edinburgh}
  \streetaddress{Informatics Forum, 10 Crichton Street}
  \city{Edinburgh}
  \postcode{EH8 9AB}
  \country{Scotland}
}
\author{Zoubin Ghahramani}
\email{zoubin@eng.cam.ac.uk}
\affiliation{
  \department{Department of Engineering}
  \institution{University of Cambridge}
  \streetaddress{Trumpington Street}
  \city{Cambridge}
  \postcode{CB2 1PZ}
  \country{England}
}
\affiliation{
  \institution{Uber AI Labs}
  \city{San Francisco}
  \country{California, USA}
}

\newcommand\newshortauthors{
\'Scibior, Kammar, V\'ak\'ar, Staton, Yang, Cai, Ostermann, Moss, Heunen, and Ghahramani
}
 \authorsaddresses{}
\begin{abstract}
  We present a modular semantic account of Bayesian inference algorithms for
  probabilistic programming languages, as used in data
  science and machine learning. Sophisticated inference algorithms are
  often explained in terms of composition of smaller parts. However,
  neither their theoretical justification nor their implementation
  reflects this modularity. We show how to conceptualise and analyse
  such inference algorithms as manipulating intermediate
  representations of probabilistic programs using higher-order
  functions and inductive types, and their denotational semantics.

  Semantic accounts of continuous distributions use measurable
  spaces. However, our use of higher-order functions presents a
  substantial technical difficulty: it is impossible to define a
  measurable space structure over the collection of measurable
  functions between arbitrary measurable spaces that is compatible
  with standard operations on those functions, such as function
  application. We overcome this difficulty using quasi-Borel spaces,
  a recently proposed mathematical structure that supports both
  function spaces and continuous distributions.

  We define a  class of semantic structures for representing
  probabilistic programs, and semantic validity criteria for
  transformations of these representations in terms of distribution
  preservation. We develop a collection of building blocks for
  composing representations. We use these building blocks to validate
  common inference algorithms such as Sequential Monte Carlo and
  Markov Chain Monte Carlo. To emphasize the connection between the semantic manipulation and its traditional measure theoretic origins, we use Kock's
  synthetic measure theory. We demonstrate its usefulness by proving a
  quasi-Borel counterpart to the Metropolis-Hastings-Green
  theorem.
\end{abstract}


\maketitle
\renewcommand{\shortauthors}{\newshortauthors}
\begin{acks}

  Supported by a Royal Society University Research Fellowship,
  Institute for Information \& Communications Technology Promotion
  (IITP) grant funded by the Korea government (MSIP) No.~R0190-16-2011
  `Development of Vulnerability Discovery Technologies for IoT
  Software Security', Engineering and Physical Sciences Research
  Council (ESPRC) Early Career Fellowship EP/L002388/1 `Combining
  viewpoints in quantum theory', an EPSRC studentship and grants
  EP/N007387/1 `Quantum computation as a programming language' and
  EP/M023974/1 `Compositional higher-order model checking: logics,
  models, and algorithms', a Balliol College Oxford Career Development
  Fellowship, and a University College Oxford Junior Research
  Fellowship. We would like to thank Samson Abramsky, Thorsten
  Altenkirch, Bob Coecke, Mathieu Huot, Radha Jagadeesan, Dexter
  Kozen, Paul B.~Levy, and the anonymous reviewers for fruitful
  discussions and suggestions.
\end{acks}

\section{Introduction}
One of the key challenges in Bayesian data analysis is to develop or find
an efficient algorithm for estimating the
posterior distribution of a probabilistic model with respect to a given
data set. This posterior distribution combines prior knowledge encoded
in the model and information present in the data set consistently according
to the rules of probability theory, but its mathematical definition often
involves integration or summation over a large index set
and does not yield to an efficient computation strategy immediately.
A data scientist typically has to make one of the suboptimal decisions:
she has to consult a large body of specialised
research in order to pick an algorithm suitable for her model, or
to change the model so that it falls into one of those cases with efficient known
algorithms for posterior inference, or to face the challenge directly
by developing a new algorithm for herself.

Recent probabilistic programming languages aim to
resolve this dilemma. They include constructs for describing
probability distributions and conditioning, and enable data
scientists to express sophisticated probabilistic models as programs.
More importantly, they come with the implementation of multiple algorithms for
performing posterior inference for models and data sets
expressed in the languages. The grand vision is that by using these languages,
a data scientist no longer has to worry about the choice or design of
such an inference algorithm but focuses on the design of an appropriate model, instead.

In this paper, we provide a denotational validation of inference algorithms
for higher-order probabilistic programming languages, such as Church~\cite{goodman_uai_2008},
Anglican~\cite{wood-aistats-2014} and Venture~\cite{Mansinghka-venture14}. The correctness of these
algorithms is subtle. The early version of the lightweight Metropolis-Hastings algorithm
had a bug because of an incorrect acceptance ratio \cite{Wingate2011}. The correctness
often relies on intricate interplay between facts from probability theory
and those from programming language theory. Moreover, correctness typically requires
stronger results from probability theory than those used for the usual
 $\reals^n$ case in the machine-learning community
(e.g., Green's measure-theoretic justification of Markov
Chain Monte Carlo rather than the usual one for $\reals^n$
based on density functions).

Our starting point is the body of existing results on validating inference algorithms for probabilistic
programs~\cite{HurNRS15,BorgstromLGS-icfp16}.
Those earlier results tend to be based on operational semantics, and often (not always) focus on first-order programs.
By working in a modular way with monads, denotational semantics and higher-order functions, we are able to validate
sophisticated inference algorithms, such as resample-move Sequential Monte Carlo~\cite{doucet-johansen:smc-tutorial11},
that are complex yet modular, being composed of smaller reusable components, by combining our semantic
analysis of these components.

The probabilistic programming language considered in the paper includes
continuous distributions, which means that semantic accounts
of them or their inference algorithms need to use measure theory and Lebesgue integration.
Meanwhile, our semantic account uses a meta-language
with higher-order functions for specifying and interpreting intermediate
representations of probabilistic programs that are manipulated by
components of inference algorithms.
Such higher-order functions let us achieve modularity and handle
higher-order functions in the target probabilistic programming language.
These two features cause a tension because it is impossible to define a
measurable space structure over the collection of measurable
functions between arbitrary measurable spaces that is compatible
with standard operations on those functions, such as function
application. We resolve the tension using quasi-Borel spaces~\cite{HeunenKSY-lics17},
a recently proposed mathematical structure that supports both
function spaces and continuous distributions.

We define a semantic class of structures for
various intermediate representations of probabilistic programs,
and semantic validity criteria for
transformations of these representations in terms of distribution
preservation. We develop a collection of building blocks for
composing representations. We use these building blocks to validate
common inference algorithms such as Sequential Monte Carlo and
Markov Chain Monte Carlo. To emphasize the connection between the semantic manipulation and its traditional measure theoretic origins, we use Kock's
synthetic measure theory. We demonstrate its usefulness by proving a
quasi-Borel counterpart to the Metropolis-Hastings-Green
theorem.

To ease the presentation, we proceed in two steps. First, we present
our development in the discrete setting, where the set-theoretic
account is simpler and more accessible. Then, after
developing an appropriate mathematical toolbox, we transfer this
account to the continuous case. Inference in the continuous setting,
while conceptually very similar to the discrete case, is inseparable
from our development. The semantic foundation for continuous
distributions over higher-order functions has been very problematic in
the past. The fact that our approach \emph{does} generalise to the
continuous case, and does so smoothly, is one of our significant
contributions, only brought about through the careful combination of
quasi-Borel spaces, synthetic measure theory, the meta-language, and
the inference building blocks.

The rest of the paper is structured as follows.
\secref{calculus} presents a core calculus, our metalanguage, with its type system and
set-theoretic denotational semantics.
\secref{discrete} presents the core ideas of our development in a
simpler discrete set-theoretic setting.
\secref{preliminaries} reviews the mathematical concepts required for
dealing with continuous distributions.
\secref{continuous} presents representations and transformations for
continuous distributions.
\secref{SMC} decomposes the common Sequential Monte Carlo inference
algorithm into simpler inference representations and
transformations.
\secref{MHG} similarly decomposes the general Trace
Markov Chain Monte Carlo algorithm.
\secref{conclusion} concludes.
Basic results in synthetic measure theory are listed for the reader's convenience in Appendix \ref{sec:kock-equations}.

 \section{The core calculus}\seclabel{calculus}
We use a variant of the simply-typed $\lambda$-calculus with sums and
inductive types, base types and constructors, primitives, and
primitive recursion, but without effects. We also use monad-like
constructs in the spirit of \citepos*{Moggi89}{computational
  $\lambda$-calculus}.  The core calculus is very simple, and at
places we need an inherently semantic treatment, which the core
calculus alone cannot express. In those cases, we resort directly to
the semantic structures, sets or spaces. However, the calculus still
serves a very important purpose: every type and function expressed in
it denote well-formed objects and well-formed morphisms. In the
continuous case, using this calculus yields correct-by-construction
quasi-Borel spaces and their morphisms, avoiding a tedious and
error-prone manual verification. Using the core calculus also brings
our theoretical development closer to potential implementations in
functional languages.

\subsection{Syntax}
\begin{figure}
  \begin{syntax}
  \ty,\ty[2],\ty[3] & \gdefinedby &           &\syncat{types}   \\
      &    & \tvar     &\synname{positive variable}\\
      &\gor& \Variant{
                \Inj{\Cns_1}{\ty_1}
                \vor \ldots \vor
                \Inj{\Cns_n}{\ty_n}
              }        &\synname{variant}          \\
      &\gor& \Unit ~\gor~ \ty \t* \ty[2] & \synname{finite product} \\
    &\gor& \rec \tvar \ty              & \synname{inductive type}       \\
\end{syntax}
~
\begin{syntax}
  \\
      &\gor& \ty \To \ty[2]              & \synname{function}      \\
      &\gor& \Type     & \synname{base}\\
      &\gor& \Functor \ty&\synname{base constructors}\\
  \multicolumn{3}{@{}l@{}}{\ctx \definedby{}  x_1 : \ty_1, \ldots, x_n : \ty_n }& \syncat{variable contexts}
\end{syntax}

  \begin{tabular}{cc}
{\begin{syntax}
  \trm, \trm[2], \trm[3] & \gdefinedby & & \syncat{terms}                          \\
  &    & \var                          & \synname{variable}                        \\
  &\gor&\tInj\ty\Cns\trm               & \synname{variant constructor}             \\
  &\gor& \tUnit
  \ \gor\ \tpair{\trm}{\trm[2]}        & \synname{nullary and binary tuples}       \\
  &\gor& \troll\ty                     & \synname{iso-inductive constructor}       \\
  &\gor& \expfun \var\ty \trm          & \synname{function abstraction} \\
  &\gor& \mathrlap{\begin{array}[t]{@{}r@{\,}l@{}l@{}}\vmatch {\trm  \\}
                {
                      &\Inj{\Cns_1}{\var_1}\To{\trm[2]_1}
                \vor \cdots
                \vor  \Inj{\Cns_n}{\var_n}\To{\trm[2]_n}
                }\end{array}}           & \synname{pattern matching: variants}
\end{syntax}
}&
{\begin{syntax}&\hphantom{\gdefinedby}\\
  &\gor& \mathrlap{\begin{array}[t]{@{}r@{\,}l@{}l@{}}\pmatch
           {\trm\\}
           {&\var[1]}{\var[2]}
           {\trm[2]}
         \end{array}}
                                       & \synname{binary products} \\
  &\gor& \mathrlap{\begin{array}[t]{@{}r@{\,}l@{}l@{}}
                   \rmatch[{&}]
                   {\trm\\}
                   \var{\trm[2]}
                \end{array}}
                                       & \synname{inductive types} \\

  &\gor& \tfold\ty\trm                 & \synname{inductive recursion}             \\
  &\gor& \trm\, \trm[2]                & \synname{function application}            \\
  &\gor& \tcts                         & \synname{primitive}
\end{syntax}
}\end{tabular}
   \caption{Core calculus types (top) and terms (bottom)}\figlabel{types}\figlabel{terms}
\end{figure}

\figref{types}~(top) presents the types of our core calculus. To support
inductive types, we include type variables, taken from a countable set
ranged over by $\tvar, \tvar[2], \tvar[3], \ldots$. Our kind system
will later ensure these type variables are \emph{strictly positive}:
they can only appear free covariantly --- to the right of a
function type. Variant types use constructor labels taken from a
countable set ranged over by $\Cns, \Cns_1, \Cns_2, \ldots$. Variant
types are in fact partial functions with a finite domain from the set
of constructor labels to the set of types. When $\ty[2]$ is a variant
type, we write $(\Inj\Cns \ty) \in \ty[2]$ for the assertion that
$\ty[2]$ assigns the type $\ty$ to $\Cns$. We include the standard
unit type, binary products, and function types.  We include unary
uninterpreted base types and constructors.  While we use a list syntax
for variable contexts $\ctx$, they are in fact partial functions with
a finite domain from the countable set of variables, ranged over by
$\var, \var[2], \var[3], \ldots$, to the set of types.

We desugar stand-alone labels in a variant type
$\Variant{\cdots \vor \Inj\Cns \vor \cdots}$ to the unit type
$\Variant{\cdots \vor \Inj\Cns\tUnit \vor \cdots}$. We also desugar
seemingly-recursive type declarations
$\ty \definedby \subst{\ty[2]}{\sfor \tvar\ty}$ to
$\ty \definedby \rec\tvar\ty[2]$.

\begin{example}\examplelabel{types}
  The type of booleans is given by \(
  \boolty \definedby \Variant{\tru \vor \fls}
  \).
  The type of natural numbers is given by \(
  \naturals
  \definedby
                  \Variant{\Inj[]{Zero}
                  \vor     \Inj{Succ}\naturals}
  \)
  desugaring to \(
  \naturals
  \definedby \rec \tvar
                  \Variant{\Inj[]{Zero}
                  \vor     \Inj{Succ}\tvar}
  \).
  The type of $\alpha$-lists is given by
  \(
  \List\tvar
  \definedby
  \Variant{
    \Inj[]{Nil}
    \vor
    \Inj{Cons}{\tvar\t* \List\tvar}
  }
  \), desugaring to
  \(
  \List\tvar
  \definedby
  \rec{\tvar[2]}\Variant{
    \Inj[]{Nil}
    \vor
    \Inj{Cons}{\tvar\t* \tvar[2]}
  }
  \).
\end{example}

Base types and constructors allow us to include semantic type
declarations into our calculus. For example, we will always include
the following base types:
\begin{center}
\vspace{-.4cm}
\begin{tabular}{@{}*3{l}@{}}
\textbullet\ $\I\,\,  $: unit interval $[0,1]$; &
\textbullet\ $\ereals $: extended real line $[-\infty, \infty]$; &
\textbullet\ $\ereals+$: non-negative extended reals \\
\textbullet\ $\reals  $: real line $(-\infty, \infty)$; &
\textbullet\ $\reals+ $: non-negative reals $[0, \infty)$; and &
\hspace{.75cm}$[0, \infty]$.
\end{tabular}
\end{center}
In addition, once we define a type constructor such as $\List\tvar$,
we will later reuse it as a base type constructor $\List\ty$,
effectively working in an extended calculus. Thus we are working with
a family of calculi, extending the base signature with each type
definition in our development.

\figref{terms}~(bottom) presents the terms in our core calculus.  Variant
constructor terms $\tInj\ty\Cns\trm$ are annotated with their variant
type $\ty$ to avoid label clashes. The tupling constructors are
standard. We use \emph{iso-inductive} types: construction of inductive
types requires an explicit rolling of the inductive definition such as
$\troll\naturals{(\Inj[]{Zero}\tUnit)}$. Variable binding in function
abstraction is \emph{intrinsically typed} in standard Church-style. We
include standard pattern matching constructs for variants, binary
products, and inductive types. We include a structural recursion
construct $\tfold\ty$ for every inductive type $\ty$. Function
application is standard, as is the inclusion of primitives.

To ease the construction of terms, we use the standard syntactic sugar
(e.g. $\letin\var \trm{\trm[2]}$ for $(\fun \var\trm)\trm[2]$,
$\tif{\,}{\,}{}$ for pattern matching booleans),
informally elide types from the terms, elide $\roll$ing/unrolling inductive types,
and informally use nested pattern matching.

\begin{example}\examplelabel{terms}
  For $\List \ty =
  \rec{\tvar}\Variant{
    \Inj[]{Nil}
    \vor
    \Inj{Cons}{\ty\t* \tvar}
  }$, we can express standard list manipulation:
  \begin{align*}
    x :: x_s &= \Inj[]{Cons}{\tpair x{x_s}}
    &
    \foldlist a f &= \tfold{\List \ty} \pfun {
          \Inj[]{{Nil}}             \To a
          \vor{}\Inj{{Cons}}{\tpair{x}{b}\ } \To f(x, b)
    }
    \\
    x_s \++ y_s &= \foldlist {y_s}{(::)}\ x_s
    &
      \tmap f\, x_s
      &=
  \foldlist {\lst-} {(\pfun{\tpair{y}{y_s}\To\tpair{f(y)}{y_s}})}
  \end{align*}
  where we abbreviate $\lst{a_1, \ldots ,a_n}$ to
  \(
  \Inj{{Cons}}{\tpair{a_1}{\ldots,\Inj{{Cons}}{\tpair{a_n}{\Inj[]{{Nil}}}\ldots}}}
  \).

\end{example}

\subsection{Type system}
\begin{figure}
  \[
\begin{array}{@{}c@{}}
  \inferrule{
    ~
  }{
    \Dkinf \tvar
  }
  (\tvar \in \Tyctx)
  \quad
  \inferrule{
    \text{for all $1 \leq i \leq n$: }
    \Dkinf {\ty_i}
  }{
    \Dkinf{\Variant{
                \Inj{\Cns_1}{\ty_1}
                \vor \ldots \vor
                \Inj{\Cns_n}{\ty_n}
              }}
  }
  \quad
  \inferrule{
    ~
  }{
    \Dkinf\Unit
  }
\\
  \inferrule{
    \Dkinf{\ty}
    \\
    \Dkinf{\ty[2]}
  }{
    \Dkinf{\ty\t*\ty[2]}
  }
\quad
  \inferrule{
    \Dkinf[, \tvar]{\ty}
  }{
    \Dkinf{\rec \tvar \ty}
  }
  \quad
  \inferrule{
    \shade{\kinf \ty : \typ}
    \\
    \Dkinf{\ty[2]}
  }{
    \Dkinf{\ty \To \ty[2]}
  }
\\
  \inferrule{
    ~
  }{
    \Dkinf{\Type}
  }
\quad
  \inferrule{
    \Dkinf \ty
  }{
    \Dkinf{\Functor \ty}
  }
\quad
  \inferrule{
    \text{for all $(x : \ty) \in \ctx$: }
    \kinf \ty : \typ
  }{
    \kinf{\ctx} : \context
  }
\end{array}
\]
   \caption{Core calculus kind system}\figlabel{kind system}
\end{figure}

To ensure the well-formedness of types, which involve type variables,
we use a simple kind system, presented in \figref{kind system}. Each
kinding judgement $\Dkinf\ty$ asserts that a given type $\ty$ is
well-formed in the \emph{type variable context} $\Tyctx$, which is
finite set of type variables.

The kinding judgements are standard. All type variables must be bound
by the enclosing context, or by an inductive type binder.  The
contravariant position in the function type $\ty[1] \To \ty[2]$ must
contain a \emph{closed} type, ensuring that free type variables can
only appear in strictly positive positions. Variable contexts $\ctx$
must only assign closed types.

\begin{example}
  The types from \exampleref{types} are well-kinded:
    $\kinf \boolty, \naturals, \List\tvar : \typ$.
\end{example}
We define capture avoiding substitution of types for type variables in
the standard way, which obeys the usual structural
properties. Henceforth we consider only well-formed types in context,
leaving the context implicit wherever possible, and gloss over issues
of alpha-convertibility of bound type variables.

To type terms, we assume each primitive $\tcts$ has a well-formed type
$\kinf \ty_{\tcts} : \typ$ associated with it. \figref{type system}
presents the resulting type system. Each typing judgement
$\Ginf\trm\ty$ asserts that a given term $\trm$ is well-typed
with the well-formed closed type $\kinf \ty : \typ$ in the
variable context $\kinf \ctx : \context$.

\begin{figure}
  \[
\begin{array}{@{}c@{}}
  \inferrule{
    ~
  }{
    \Ginf \var\ty
  }((\var : \ty) \in \ctx)
  \quad
  \inferrule{
    \Ginf\trm{\ty_i}
  }{
    \Ginf{\tInj\ty{\Cns_i}\trm}{\ty}
  }((\Inj{\Cns_i}\ty_i) \in \ty)
\quad
  \inferrule{
    ~
  }{
    \Ginf\tUnit\Unit
  }
\\
  \inferrule{
    \Ginf {\trm[1]}{\ty[1]}
    \\
    \Ginf {\trm[2]}{\ty[2]}
  }{
    \Ginf{\tpair{\trm[1]}{\trm[2]}}{\ty[1] \t* \ty[2]}
  }
\quad
  \inferrule{
    ~
  }{
    \Ginf{\troll\ty}{\parent{\subst{\ty[2]}{\sfor{\tvar}{\ty}}}\To\ty}
  }(\ty = \rec\tvar\ty[2])
\quad
  \inferrule{
    \Ginf[, \var : \ty]{\trm}{\ty[2]}
  }{
    \Ginf{\expfun \var\ty\trm}{\ty\To\ty[2]}
  }
\\
  \inferrule{
    \Ginf\trm{\Variant{
                \Inj{\Cns_1}{\ty_1}
                \vor \ldots \vor
                \Inj{\Cns_n}{\ty_n}}}
    \\
    \text{for each $1 \leq i \leq n$: }
    \Ginf[, \var_i : \ty_i]{\trm[2]_i}{\ty}
  }{
    \Ginf{\vmatch \trm
                {\begin{array}[t]{@{}l@{\,}l@{}l@{}}
                    \Inj{\Cns_1}{\var_1}\To{\trm[2]_1}
                    \vor\cdots
                    \vor\Inj{\Cns_n}{\var_n}&\To{\trm[2]_n}
    }}
    \ty
                  \end{array}
  }
\\
  \inferrule{
    \Ginf{\trm}{\ty[2]\t*\ty[3]}
    \\
    \Ginf[,{\var[1] : \ty[2], \var[2] : \ty[3]}]{\trm[2]}\ty
  }{
    \Ginf{\pmatch
           \trm
           {\var[1]}{\var[2]}
           {\trm[2]}}{\ty}
  }
  \quad
  \inferrule{
    \Ginf{\trm}{\rec\tvar\ty[2]}
    \\
    \Ginf[,{\var : \subst{\ty[2]}{\sfor\tvar{\rec\tvar\ty[2]}}}]{\trm[2]}\ty
  }{
    \Ginf{\rmatch
           \trm
           {\var}
           {\trm[2]}}{\ty}
  }
\\

  \inferrule{
    \Ginf\trm{\parent{\subst{\ty[2]}{\sfor\tvar{\ty[3]}}}\To \ty[3]}
  }{
    \Ginf{\tfold\ty\trm}{\ty \To \ty[3]}
  }(\ty = \rec\tvar\ty[2])
  \quad
  \inferrule{
    \Ginf{\trm}{\ty[2]\To\ty}
    \\
    \Ginf{\trm[2]}{\ty[2]}
  }{
    \Ginf{\trm\, \trm[2]}{\ty}
  }
\quad
  \inferrule{
      ~
    }{
      \Ginf{\tcts}{\ty_{\tcts}}
    }
\end{array}
\]
   \caption{Core calculus type system}\figlabel{type system}
\end{figure}

The rules are standard. By design, every term has at most one type in
a given context.

\begin{example}
  Once desugared, the list manipulation terms from \exampleref{terms}
  have types:
  \begin{align*}
    (::) &: \ty \t* \List \ty \To \List \ty
    &
    \foldlist &: \ty[2] \t* (\ty\t*\ty[2] \To \ty[2])
                         \t* \List\ty
             \To \ty[2]
    \\
    \tmap &: (\ty \To \ty[2])\To (\List \ty \To \List \ty[2])
    &
    (\++) &: (\List \ty) \t* (\List \ty) \To \List \ty
  \end{align*}
\end{example}

\subsection{Primitive recursion}

As is
well-known~\cite{hutton:a-tutorial-on-the-universality-and-expressiveness-of-fold,
  geuvers2007iteration}, structural recursion on inductive types
allows us to express primitive recursion.  By `primitive recursion', we
mean recursing through values of an inductive type $\rec\tvar{\ty[2]}$
using a term of the form:
$\ctx, k : \subst{\ty[2]}{\sfor\tvar{(\rec\tvar{\ty[2]})\t*\ty[3]}}
\tinf \trm[1] : \ty[3]$
with the intention that $\trm[1]$ can use either arbitrary (total) processing
on the sub-structures of its input $k$, or make a primitive recursive
call to itself with a sub-structure. In order to desugar such a term
into a function of type $\ty \t* (\rec\tvar\ty[2]) \To \ty[3]$, we use terms of the following type, defined by induction on types:
\[
  \projection_{\tvar.\ty[2],\ty[3]} :
  \subst{\ty[2]}{\sfor\tvar{(\rec\tvar{\ty[2]})\t*\ty[3]}} \to
  \subst{\ty[2]}{\sfor\tvar{\rec\tvar{\ty[2]}}}
\]
and interpret the primitive recursive declaration $t$ embodied by:
\[
  \ctx, \var : \rec\tvar\ty[2]  \tinf
    \begin{array}[t]{@{}*3{l@{}}}
      \pmatch
    {\tfold{(\rec\tvar\ty[2])}
      {}&
        \parent{\expfun k{\subst{\ty[2]}{\sfor\tvar{(\rec\tvar{\ty[2]})\t*\ty[3]}}}
	\,\tpair {\roll \projection_{\tvar.\ty[2],\ty[3]}k}{t}}
      \var\\\!
  } {\_}{r}{r} : \ty[2]
    \end{array}
\]
This translation is global in nature: the structure of the term
$\projection$ depends on the type of $t$. Thus, it does not constitute
a \emph{macro}
translation~\cite{felleisen:on-the-expressive-power-of-programming-languages}.
With this point in mind, we will allow ourselves to use primitive
recursive definitions.

\begin{example}\examplelabel{aggr intro}
  We define a function
  $\aggr : \List(\reals+\t*X) \To \List(\reals+\t*X)$ which takes a
  list of weighted values and aggregates all the weights based on
  their values. We make use of the auxiliary function
  $\add : (\reals+\t* X) \t* \List (\reals+ \t*X) \To \List
  (\reals+ \t*X)$, which adds a weighted value to an already
  aggregated list. We define $\add$ by primitive recursion:
  \[
    \add \tpair{\tpair sa}{x_s} \definedby
       \vmatch {x_s} {\begin{array}[t]{@{}l@{\To}l}
           \lst{\,}         \;& \;\lst{\tpair sa} \mspace{175mu}\scomment{new entry}\\
           \tpair rx :: x_s \;& \;\begin{array}[t]{@{}l@{\,}l@{\qquad}l}
             \tif
                 {&x = a\\}
                 {&\tpair {s + r}a :: x_s          &\scomment{accumulate}\\}
                 {&\tpair rx :: \add\tpair{\tpair sa}{x_s}}\} &\scomment{recurse}\gobble
       }           \end{array}

      \end{array}
  \]
  and set $\aggr \definedby \foldlist {\lst-}\add$. This example makes
  use of an equality predicate between $X$ elements, restricting its
  applicability.
\end{example}

\subsection{Denotational semantics}
We give a set-theoretic semantics to the calculus. In such
set-theoretic semantics, types-in-context $\Dkinf\ty$ are interpreted
as functors $\sem{\ty} : \Set^{\Tyctx} \to \Set$, i.e., $\sem{\ty}$
assigns a set $\sem{\ty}\seq[\tvar \in \Tyctx]{X_{\tvar}}$ for every
$\Tyctx$-indexed tuple of sets, and a function
\[ \sem\ty\seq[\tvar \in \Tyctx]{f_{\tvar} : X_{\tvar} \to Y_{\tvar}}
  : \sem\ty\seq{X_{\tvar}} \to \sem\ty\seq{Y_{\tvar}}
\]
for every $\Tyctx$-indexed tuple of functions between the sets with
corresponding index, and this assignment preserves composition and
identities.

In order to interpret iso-inductive types $\rec\tvar\ty$, we need
canonical isomorphisms between the sets
$\sem{\ty} (\sem{\rec\tvar\ty}) \isomorphic \sem{\rec\tvar\ty}$.
We will do this in a standard way, by interpreting $\sem{\rec\tvar \ty}$ as the initial algebra
for the functor $\sem{\ty}:[\Set^\Delta\to \Set]\to [\Set^\Delta\to \Set]$.
This means that for every functor $A:\Set^\Delta\to \Set$
with a natural family of functions $\{a_{X}:(\sem{\ty} A)(X)\to A(X)\}_{X\in\Set^\Delta}$,
there is a canonical natural family of functions
$\{\algfold_X : \sem{\rec\tvar\ty}(X) \to A(X)\}_{X\in\Set^\Delta}$.

A technical requirement is needed to ensure that this initial algebra exists:
we fix a regular cardinal $\kappa$, and
demand that each type denotes a $\kappa$-\emph{ranked} functor
(ranked functor for short),
that is, that it denotes a functor that preserves $\kappa$-filtered
colimits\footnote{
We do not use simpler classes of functors, such as \emph{polynomial
  functors} or \emph{containers}, as they are not closed under
subfunctors, given by subsets in the discrete case and subspaces in
the continuous case, which we need in the sequel.
}.  The
$\kappa$-ranked functors are closed under composition, products, sums,
and initial algebras.  Initial algebras for $\kappa$-ranked functors
on locally presentable categories always exist, because they can be
built in an iterative way by transfinite induction (see
e.g.~\cite{kelly-transfinite}).

\subsubsection{Set-theoretic interpretation}\subsubseclabel{interpretation}
To interpret types, we assume a given interpretation $\Base-$ of
the base types $\Type$ as sets $\Base{\Type}$ and of base type
constructors $\Functor$ as ranked functors
$\Base\Functor : \Set \to \Set$. We then interpret each well-formed
type in context $\Dkinf\ty$ as a ranked functor
$\sem{\ty} : \Set^{\Tyctx} \to \Set$, as depicted in
\figref{sem types}.

\begin{figure}
  \begin{gather*}
  \sem{\alpha}\tenv \definedby \tenv(\alpha)
  \qquad
  \sem{\Variant{
                \Inj{\Cns_1}{\ty_1}
                \vor \ldots \vor
                \Inj{\Cns_n}{\ty_n}
  }}\tenv
  \definedby
  \sum_{i = 1}^n \sem{\ty_i}\tenv
  \qquad
  \sem{\Unit} \definedby %\terminal
  \qquad
   \sem{\Functor \ty}\tenv \definedby \Base\Functor(\sem\ty\tenv)
  \\[-12pt]
  \sem{\rec \tvar \ty}\tenv \definedby \rec X{\sem\ty}\subst\tenv{\sfor \tvar X}
  \qquad
  \sem{\ty \To \ty[2]}\tenv \definedby \parent{\sem{\ty[2]}\tenv}^{\shade{\sem{\ty}\seq{}}}
  \qquad
  \sem{\Type}\tenv \definedby \Base\Type
\end{gather*}
   \caption{Core calculus type-level semantics}\figlabel{sem types}
\end{figure}

In this definition, the parameter $d$ may be either a tuple of sets or
functions. When interpreting type variables, we write $d(\alpha)$ for
the $\alpha$-indexed component of $d$. The interpretation of simple
types uses disjoint unions, singletons, finite products, and
exponentials, i.e.~the bi-cartesian closed structure of $\Set$.  We
interpret inductive types $\sem{\rec \tvar\ty}\tenv$ using the initial
algebra for the ranked functor  $\fun X \sem\ty\subst\tenv{\sfor \tvar X} : \Set \to \Set$.  In the
semantics of the function type $\ty[1]\To\ty[2]$, the exponential
makes no use of the functor's arguments, and relies on the fact that
all type variables are strictly positive. We use the given
interpretation of base types and type constructors to interpret them.

\begin{lemma}\lemmalabel{well type sem}
  The semantics of types is well-defined: every well-formed type
  $\Dkinf \ty$ denotes a ranked functor
  $\sem\ty : \Set^\Delta \to \Set$. In particular, every closed type
  denotes a set.
\end{lemma}

The proof is by induction on the kinding judgements, using well-known
properties of $\Set$.

We will always interpret the base types $\I$, $\reals$, etc.~by the
sets they represent.
\begin{example}
  We calculate the denotations of the types from
  \exampleref{types}. Booleans denote a two-element set
  $\sem\boolty = \set{\fls, \tru}$, and the natural numbers denote the
  set of natural numbers $\sem\naturals = \naturals$. By
  \lemmaref{well type sem}, $\sem{\List}$ denotes a ranked functor
  $\List : \Set \to \Set$, and this functor is given by the set of
  sequences of $X$-elements
  $\List X \definedby \Union_{n \in \naturals}X^n$.
\end{example}

Beyond establishing the well-definedness of the semantic
interpretation, \lemmaref{well type sem} equips us with syntactic
means to define ranked functors. Once defined, we can add these
functors to our collection of base types (in an extended instance of
the core calculus). In the sequel, we will often restrict a given
ranked functor $F : \Set \to \Set$ by specifying a subset
$G X \subset FX$. Doing so is analogous to imposing an
\emph{invariant} on a datatype. The subsets $GX$ form a
\emph{subfunctor} $G \subset F$ precisely if they are closed under the
functorial action of $F$, i.e., for every function $f : X \to Y$ and
$a \in GX$, $Ff(a) \in GY$.

\begin{lemma}
  Subfunctors of ranked functors over $\Set$ are ranked.
\end{lemma}

We can prove this lemma directly, but it also follows from a
higher-level argument using the commutation of finite limits and
$\kappa$-directed colimits in $\Set$.

\subsection{Monadic programming}\subseclabel{monad interfaces}
In the sequel, we will be working with types that support a monadic
programming style. More precisely, a \emph{monadic interface}
$\monad T$ consists of a triple
$\monad T = \triple* T{\return^{\monad T}}{\bind^{\monad T}}$ where:
$T$ assigns to each set $X$ a set $TX$; $\return^{\monad T}$ assigns
to each set $X$ a function $\return_X^{\monad T} : X \to TX$; and
$\bind^{\monad T}$ assigns to each pair of sets $X$ and $Y$ a function
$\bind_{X, Y}^{\monad T}{} : TX\times (TY)^X \to TY$. We borrow
Haskell's type-class syntax to define such interfaces. As an example,
\figref{instance} defines a monadic interface over $\List$.

Each such monadic interface $\monad T$ allows us to use standard
do-notation summarised in \figref{do}. Though simple in principle, we
must take care when treating this notation as syntactic sugar, as
choosing the appropriate function $\return_X$ or $\bind_{X,Y}$ at each
desugaring step must take typing information into account. When we use
do-notation in the sequel, we ensure that such choices can be
disambiguated. Finally, we will delimit our use of do-notation to
within a \emph{do-block} $\expdomon{\monad T}{\ldots}$, omitting the
monadic interface $\monad T$ or the entire delimiter when either is
clear from the context.

\begin{figure}
\begin{minipage}[t]{.5\linewidth}
  \(
  \Monad{\List}{
    \sreturn x &{}= \lst x
  }{
    x_s \bind f  &{}=
                        \foldlist {\lst-}
                                  \parent{\fun{\tpair x{y_s}}
                                  f(x) \++ y_s}\ x_s
  \\\\}
\)
 \subcaption{Declaring monadic interfaces}\figlabel{instance}
\end{minipage}
\begin{minipage}[t]{.5\linewidth}
\center
\begin{sugar}
  \var  \<- \trm; \trm[2]  & \trm \bind \fun \var\trm[2] \\
  \sreturn \trm                   & \return^{\monad T} \trm           \\
  \trm; \trm[2]           & \_ \<- \trm; \trm[2]                      \\
\end{sugar}
\subcaption{Haskell's do-notation}\figlabel{do}
\end{minipage}
\caption{Monadic programming notation}
\end{figure}

Importantly, we do not insist that a monadic interface satisfies
the monad associativity and unit laws:
\(
  (\return x) \bind f = f(x)
  \), \(
  a \bind \return x = a
  \), and \(
  (a \bind f) \bind g = a \bind (\fun x (f\ x \bind g))
  \).
 \section{Discrete inference}\seclabel{discrete}
We can now lay-out the core ideas in the simpler, set-theoretic case:
a semantic structure for higher-order (discrete) probabilistic
programs, intermediate representations of these programs for the
purpose of inference, valid transformations between these
representations, and modular building blocks for creating new
representations and transformations from existing ones. For simplicity,
we consider representations and transformations from simple rather naive
inference algorithms only in this section. In \secref{SMC} and
\secref{MHG}, we show how the core ideas here apply to
advanced algorithms when aided with further technical developments.

\subsection{The mass function monad}
For our purposes, probabilistic programming languages contain
standard control-flow mechanisms and data types, such as our core
calculus, together with \emph{probabilistic choice} and
\emph{conditioning} operations. In the discrete case, these are given
by two effectful operations:
\begin{mathpar}
  \inferrule{
    ~
  }{
    \Ginf{_{\mathrm{comp}} \sflip} \boolty
  }

  \inferrule{
    \Ginf {\trm} \reals+
  }{
    \Ginf {_{\mathrm{comp}}\sscore\trm} \Unit
  }
\end{mathpar}
In Bayesian probabilistic programming, we think of $\sflip$ as drawing from a (uniform) prior distribution on $\boolty$,
and of $\sscore$ as recording a likelihood. Typically, one calls $\sscore(f(x))$
where $f$ is a density function of a distribution, which records the likelihood of
observing data $x$ from the distribution $f$.
The score might be zero, a hard constraint: this path is impossible.
The score might be in the unit interval, the probability of a discrete observation;
but in general a likelihood function can take any positive real value.
The inference problem is to approximate the posterior distribution, from the unnormalized posterior defined by the program, combining a prior and likelihood.

To give a set-theoretic semantic structure to such a higher-order
language with these two constructs, it suffices to give a monadic
interface $\monad T$ for which the associativity and unit laws hold,
together with two functions:
\begin{mathpar}
  \flip : \sem\Unit \to T\sem\boolty

  \score: \sem{\reals+} \to T\sem\Unit
\end{mathpar}

For the purposes of the discrete development, the following monad fits
the bill. A (finite) \emph{mass function} over a set $X$ is a function
$\mu : X \to \reals+$ for which there exists a finite set
$F \subset X$ such that $\mu$ is $0$ outside $F$:
in other words, the support set $\support \mu\definedby \set {x\in X\suchthat\mu(x)\neq 0}$ is finite.
For every set $X$, let
$\Mass X \definedby \set{ \mu : X \to \reals+ \suchthat \text{$\mu$ is
    a mass function}}$. The \emph{mass function} monad is given by:
\[
  \monad \Mass \definedby \Monad{\Mass}{
    \return \var_0\ \  &= \fun {\var} \tif {(\var = \var_0)}{1}{0}
  }{
    \mu \bind f      &= \fun y \sum_{x \in \support \mu}\mu(x)\cdot (f(x)(y))
  }
\]
and we set $\flip = \fun \_ \tfrac 12$ and
$\score r = \pfun {\tUnit \To r}$. Intuitively, values of $\Mass X$
represent unnormalized probabilistic computations of a result in $X$.
From the Bayesian perspective, the meaning of a program is the unnormalized posterior.

\begin{lemma}
  The monadic interface $\monad \Mass$ defines a ranked monad over $\Set$.
\end{lemma}

This monad is also known as the \emph{free positive cone monad}, as it
constructs the `positive fragment' of a vector space over the field of
reals with basis $X$.

\subsection{Inference representations}
The mass function semantics is accurate, but idealised: realistic
implementations cannot be expected to compute mass functions at
arbitrary types, and especially at higher-order types. Instead,
probabilistic inference engines would manipulate some representation
of the program, while maintaining its semantics.

\begin{definition}\definitionlabel{disc inf rep}
  A \emph{discrete inference representation} $\ir$ is a sextuple
  \[
    \ir = \seq{T, \return^{\ir}, \bind^{\ir},
      \flip^{\ir}, \score^{\ir}, \mean^{\ir}}
  \] consisting of:
  \begin{itemize}
  \item a monadic interface $\triple T{\return^{\ir}}{\bind^{\ir}}$;
  \item two functions $\flip^{\ir} : \terminal \to T\Two$ and
    $\score^{\ir}: \reals+ \to T\terminal$, where $\terminal :=
    \sem\Unit$, $\Two := \sem{\boolty}$; and
  \item an assignment of a \emph{meaning} function
    $\mean^{\ir}_X : TX \to \Mass X$ for every set $X$
  \end{itemize}
  such that the following laws hold for all sets $X$, $Y$, and
  $x \in X$, $a \in TX$, $r \in \reals+$, and $f : X \to TY$:
  \begin{mathpar}
    \return^{\monad\Mass}x = \mean (\return^{\ir} x)

    \mean (a\bind^{\ir} f) =
    (\mean a)\bind^{\monad \Mass} \fun x \mean(f\ x)

    \\

    \mean(\flip^{\ir})  = \flip^{\monad\Mass}

    \mean(\score^{\ir} r) = \score^{\monad\Mass} r
  \end{mathpar}
\end{definition}

As with monadic interfaces, we use a type-class notation for defining
inference representations.

\begin{example}[Discrete weighted sampler]\examplelabel{disc weighted sampler}
  Consider the type
  \[
    \Term \tvar \definedby \Variant{\Inj{Return}{(\reals+\t*\tvar)} \vor
                          \Inj{Flip}{(\Term\tvar\t* \Term\tvar)}}
  \]
  which induces a ranked functor $\Term$. The elements of $\Term X$
  are binary trees, which we call terms, whose leaves contain weighted
  values of type $X$.  \figref{disc free} presents the inference
  representation structure of the functor $\Term$.
  \begin{figure}
    \begin{minipage}{\linewidth}
    \[
    \DiscMonad{\Term}{
      \sreturn x &= \Inj{Return}\tpair 1x
    }{
      a  \bind f  &=
      \begin{array}[t]{@{}l@{}l@{}}
      \expletin{&scale}{\reals+\t*\Term X \to \Term X}
      {\mspace{75mu}\scomment{uses primitive recursion}\\&
           \fun s\pfun{
           &\Inj{Return}{\tpair rx} & \To \Inj{Return}{\tpair {s\cdot r}x}\\
      \vor &\Inj{Flip}{\tpair{k_{\fls}}{k_{\tru}}}
                                    & \To \Inj{Flip}
                                              \tpair{scale\tpair{r}{k_{\fls}}}
                                                    {scale\tpair{r}{k_{\fls}}}

           }
        \\\!}
      & \imatch a {
             \Inj{Return}{\tpair rx} & scale\tpair r{f\ x} \\
        \mathllap\vor \Inj{Flip}\tpair {k_{\fls}}{k_{\tru}}
                                & \Inj{Flip}
                              \begin{aligned}[t]
                                 \tpair{&k_{\fls}\bind f}
                                       {\
                                           \scomment{uses primitive recursion}
                                        \\&k_{\tru}\bind f}\}
                               \end{aligned}\gobble
        }
      \end{array}
    }
    {
      \sflip &= \Inj{Flip}{\tpair{\Inj{Return}{\tpair{1}\fls}}
                                 {\Inj{Return}{\tpair{1}\tru}}}
    }
    {
      \sscore r &= \Inj{Return}{\tpair r\tUnit}
    }
    {
      \mean a   &= \fold{\pfun{
          &\Inj{Return}{\tpair rx}
          &\To
          \expdomon\Mass{\sscore r; \sreturn x}\\
          \vor &\Inj{Flip}{\tpair {\mu_{\fls}}{\mu_{\tru}}}
           & \To \expdomon\Mass{
             \begin{array}[t]{@{}l@{}}
               x \<- \sflip; \\
               \tif x {\mu_{\tru}}{\mu_{\fls}}
           }
        }
             \end{array}
      }
    }
  \]
     \subcaption{Discrete weighted sampler representation}\figlabel{disc free}
    \vspace{1\baselineskip}
    \end{minipage}
    \begin{minipage}[t]{.55\linewidth}
    \(
  \DiscMonad{\Enum}
  {
    \sreturn x  &= \lst{\tpair 1x}
  }{
    x_s \bind f &=
         \begin{array}[t]{@{}l@{}l@{}l@{}}
           \expletin {&scale}{\reals+\t*\Enum X \To \Enum X}
                   {\\&\pfun {\tpair {r}{x_s} \To{} \tmap {}&\pfun {\tpair sy \To \tpair {r\cdot s}y}\\&x_s}\\\!}
           &
             \begin{array}[t]{@{}l@{}}
               \foldlist {\lst-} \\
                         {\pfun{\tpair{\tpair rx}{y_s} \To scale\tpair r{f\ x} \++ y_s}}\\
                          x_s
             \end{array}
         \end{array}
  }{
    \flip &= \lst{\tpair \half\fls, \tpair \half\tru}
  }{
    \score r &= \lst{\tpair r\tUnit}
  }{
    \mean x_s &= \fun a \qquad\scomment{
                        $\mean x_s\ a=\textstyle\sum_{\substack{\tpair rx \in x_s\\x = a}}r$ }\\
                        &&
                         \foldlist{
                           0 }
                           \parent{
                            \begin{array}{@{}l@{}}
                           \pfun{\tpair{\tpair rx}{s}
                                 \To \\{}\tif{x = a}{r + s}s}
                           \end{array}
                           }
                           \ x_s
  }
\)
     \subcaption{Discrete enumeration sampler}\figlabel{disc enum}
    \end{minipage}
    \begin{minipage}[t]{.35\linewidth}
      \(
  \DiscTransBreak {\Writer}
         {
           \slift_{\ir} a &= \expdomon\ir{\begin{aligned}[t]&x \<- a;\\& \sreturn \tpair 1x
         } \end{aligned}

         }
         {
           \sreturn_{\Writer\ir} x &= \return^{\ir} \tpair 1x
         }
         {
           a \bind_{\Writer\ir} f  &=
           \begin{aligned}[t]
             \expdomon\ir{&\tpair rx \<- a;
                        \\&\tpair sy \<- f(x);
                        \\&\sreturn \tpair{r\cdot s}y}
           \end{aligned}
         }
         {
           \sflip_{\Writer\ir}     &= \slift \sflip^{\ir}
         }
         {
           \sscore_{\Writer\ir} r            &= \return^{\ir} \tpair r\tUnit
         }
         {
           \mean_{\Writer\ir} a              &= \fun x\sum_{\tpair rx \in \support{\mean^{T}(a)}}r
         }
         {
           (\stmap \tfm)_X &= \tfm_{\reals+\t* X}
         }
  \)
       \vspace{1.2\baselineskip}
    \subcaption{Discrete weighting transformer}\figlabel{disc weighting}
    \end{minipage}
    \caption{Example inference representations (a,b) and transformers (c)}
  \end{figure}
  $\Inj{Flip}$ represents a probabilistic choice while $\Inj{Return}$
  holds the final value and the total weight for the branch.
  Thus an
  immediately returning computation is represented by a leaf with
  weight $1$.  The auxiliary function $scale$ in the definition of $\bind$ scales
  the leaves of its input term by the input weight. The function
  $\bind$ itself substitutes terms for the leaves according to its
  input function $f$, making sure the newly grafted terms are scaled
  appropriately. The probabilistic choice operation $\sflip$ constructs
  a single node with each leaf recording the probabilistic choice
  \emph{unweighted}. Conditioning records the input weight.

  The meaning function recurses over the term, replacing each node
  representing a probabilistic choice by probabilistic choice of the
  mass function monad, and reweighting the end result appropriately.

  The main step in validating the inference representation laws
  involves $\bind$: first show that composing the meaning function
  with the auxiliary function $scale$ scales the meaning of the input
  term appropriately, and then proceed by structural induction on terms.
\end{example}

The weighted sampler representation in fact forms a proper monad over
$\Set$: it is the free monad for an algebraic theory with a binary
operation $\flip$ and unary operations $\score_r$ subject to
\(
  \sflip(\sscore_r(x), \sscore_r(y)) = \sscore_r(\sflip(x, y))\text.
\)
As the mass function monad also validates these equations, the meaning
function is then the unique monad morphism from $\Term$ to $\Mass$
preserving the operations $\sflip$ and $\sscore$.

However, we emphasise that an inference representation need not form a
proper monad, and that the meaning function need not be a monad
morphism. Indeed, the $\Pop \Sam$ representation introduced in \secref{SMC}
is not a monad and most of the non-trivial inference transformations we
discuss are not monad morphisms.

The weighted sampler representation allows us to incorporate both
intensional and operational aspects into our development.  Bayesian
inference ultimately reduces a representation into probabilistic
simulation. The weighted sampler representation can thus act as an
internal representation of this simulation.  Moreover, its continuous
analogue will allow us to manipulate traces when analysing the Trace
Markov Chain Monte Carlo algorithm in \secref{MHG}.

\begin{example}[Enumeration]\examplelabel{enumeration}
  The type $\Enum \tvar \definedby \List (\reals+\t* \tvar)$ induces a
  ranked functor $\Enum$. Elements of $\Enum X$ form an enumeration of
  the mass function they represent, with the same value $x$
  potentially appearing multiple times with different weights.  Values
  not appearing in the list at all have weight $0$.

  \figref{disc enum} presents an inference representation structure
  using $\Enum$. Returning a value lists the unique non-zero point
  mass. The $\bind$ operation applies the given function to each
  element listed, scales the list appropriately and accumulates all
  intermediate lists. The choice operation enumerates both branches
  with equal probability, and conditioning inserts a scaling
  factor. The meaning function assigns to an element the sum of
  its weights. This definition uses an equality predicate.

  Establishing the inference representation laws is straightforward.
\end{example}

\subsection{Inference transformations}
We can now define the central validity criterion in our
development. We decompose Bayesian inference algorithms into smaller
transformations between inference representations. To be correct,
these transformations need to preserve the meaning of the
representation they manipulate:
\begin{definition}\deflabel{disc inf tfm}
  Let $\ir$, $\ir[2]$ be two inference representations. A
  \emph{discrete inference transformation} ${\tfm : \ir \to \ir[2]}$
  assigns to each set $X$ a function $\tfm_X : \uir[1]X \to \uir[2]X$
  satisfying $\mean^{\ir}(a) = \mean^{\ir[2]}(\tfm_X(a))$ for every $a
  \in \ir[1]X$.
\end{definition}
This validity criterion guarantees nothing beyond the preservation of
the overall mass function of our representation. The transformed
representation may not be better for inference along any axis, such as
better convergence properties or execution time. It is up to the
inference algorithm designer to convince herself of such properties by
other means: formal, empirical, or heuristic.

Some transformations change the representation type:
\begin{example}[Enumeration]
  Define a transformation:
  \(
  \tfm : \monad\Term \to \monad\Enum
  \) by:
  \[
  \tfm \definedby \pfun{
    \Inj{Return}\tpair rx
        & \To \expdomon\Enum{\sscore r; \sreturn x} \\
    \vor \Inj{Flip  }\tpair {x_s^{\fls}}
                       {x_s^{\tru}}
        & \To \expdomon\Enum{b \<- \sflip; \tif b {x_s^{\tru} }
                                                  {x_s^{\fls} }
  }}
  \]
  Straightforward calculation shows it preserves the meaning functions.
\end{example}
The last example is a special case: analogous functions form
inference transformations $\tfm_{\ir} : \monad\Term \to \ir$ for every
discrete inference representation $\ir$. To establish meaning
preservation, calculate that both $\mean^{\monad\Term}$ and
$\mean^{\ir}\compose\tfm_{\ir}$ are monad morphisms that preserve
probabilistic choice and conditioning and appeal to the initiality of
$\monad\Term$.

An inference transformation need not be natural:
\begin{example}[Aggregation]
  Recall the functions $\aggr_X : \List(\reals+\t* X) \to
  \List(\reals+\t* X)$ from \exampleref{aggr intro} which aggregate
  list elements according to their $X$ component by summing their weights. It forms an
  inference transformation $\aggr : \monad\Enum \to \monad\Enum$. The meaning preservation proof uses straightforward structural induction.  Note that $\aggr$ is not a natural transformation.
\end{example}

\subsection{Inference transformers}
We can decompose the weighted sampler representation $\monad\Term$,
which forms a monad, by transforming the \emph{discrete sampler}
representation $\DSam X \definedby \Variant{\Inj{Return} X \vor
  \Inj{Sample}({\DSam X}\t*{\DSam X})}$ with the following
\emph{writer monad transformer}
\(
\Writer \uir X \definedby \uir (\reals+\t* X)
\), i.e.~\(
\Term = \Writer \DSam
\).
Such decompositions form basic building blocks for constructing and
reasoning about more sophisticated representations.

\begin{definition}
  An \emph{inference transformer} $\IF$ is a triple
  $\triple*{\uIF}{\itmap^{\IF}}{\lift^{\IF}}$ whose components assign:
  \begin{itemize}
  \item inference representation $\uIF \ir$ to every
    inference representation $\ir$;
  \item inference transformation
    $\itmap^{\IF} \tfm : \IF\ir \to \IF\ir[2]$ to every inference
    transformation $\tfm : \ir \to \ir[2]$; and
  \item inference transformation $\lift_{\uir} : \ir \to \uIF \ir$ to
    every inference representation $\ir$.
  \end{itemize}
\end{definition}

We use type-class notation for defining inference transformers.

\begin{example}\examplelabel{disc weighting trans}\examplelabel{weighting}
  The \emph{weighting} inference transformer structure on $\Writer
  \uir X \definedby \uir (\reals+\t* X)$ is given in \figref{disc
    weighting}. We lift a representation in $\ir$ into $\Writer \ir$
  by assigning weight $1$ to it. The monadic interface uses the
  standard writer monad for the multiplication structure on $\reals+$,
  accumulating the weights as computation proceeds. We lift the
  probabilistic choice from $\ir$, but crucially we reimplement a
  \emph{new} conditioning operation using the explicitly given
  weights. The mass function meaning of a representation then
  accumulates the mass of all weights associated to a given value. We
  transform an inference transformation by picking the component of
  the appropriate type.

  It is straightforward to show that $\Writer\ir$ is an inference representation, using preservation of $\return$ and $\bind$ by the
  meaning function to reduce the proof to manipulations of weighted
  sums over $\reals+$. Establishing the validity of $\lift$ and
  $\itmap$ is straightforward.
\end{example}

The weighting transformer \emph{augments} the representation with a
new conditioning operation, but \emph{transforms} its choice operation
to the new representation. We will later see more examples of both
kinds.

\subsection{Summary}
We have introduced our three core abstractions, inference
representations, transformations, and transformers, in relation to a
mathematical semantic structure, the mass function monad. The examples
so far show that the higher-order structure in our core calculus acts
as a useful glue for manipulating and defining these abstractions. In
the continuous case, we will also use this higher-order structure to
represent computations over the real numbers.
 \section{Preliminaries}\seclabel{preliminaries}
In order to generalise this higher-order treatment to the continuous
case, we first need to review and develop the mathematical theory of
quasi-Borel spaces. Our development uses
\citepos*{kock:commutative-monads-as-a-theory-of-distributions}{synthetic
  measure theory}, which allows us to reason analogously to measure
theory. In order to present the synthetic theory, we briefly review
the required category theoretic concepts. These sections are aimed at
readers who are interested in the categorical context of our
development. Other readers may continue directly to \subsecref{qbs}.

\subsection{Category theory}
\paragraph{Basic notions.}
We assume basic familiarity with categories $\cat$, $\cat[2]$, functors
$\F, \F[2] : \cat \to \cat[2]$, and natural transformations
$\nt, \nt[2] : \F \to \F[2]$, and their theory of limits, colimits,
and adjunctions. To fix notation, a \emph{cartesian closed category}
is a category with finite products, denoted by $\terminal$, $\times$,
$\prod_{i=1}^n$, and exponentials, denoted by $X^Y$.
In this subsection, we use
the fragment of our core calculus consisting of the simply-typed
$\lambda$-calculus (with sums, if necessary) to more compactly review
the relevant concepts.

\paragraph{Monads.}A \emph{strong monad} $\monad T$ over a
cartesian closed category is a triple $\triple T\return\bind$
consisting of an assignment of an object $TX$ and a morphism
$\return_X : X \to TX$ for every object $X$, and an assignment of a
morphism $\bind_{X,Y} : TX \times (TY)^X \to TY$, satisfying the monad
laws from \subsecref{monad interfaces}.
Given a monad $\ir$, a
\emph{$\ir$-algebra} $\alg$ is a pair
$\pair*{\carrier \alg}{\bind^{\alg}}$ consisting of an object
$\carrier \alg$, called the \emph{carrier}, and an assignment of a
morphism $\bind^{\alg}_X : \carrier \alg^X \to \carrier\alg^{TX}$ to
every object $X$ satisfying
\[
  (\return x \bind^{\alg} f) = f\ x
  \quad\text{and}\quad
  ((a \bind f) \bind^{\alg} g)
  =
  a \bind (\fun x f(x) \bind^{\alg} g)
\text.
\]
The pair $\tpair {\uir X}\bind$ always forms a $\uir$-algebra called
the \emph{free} $\ir$-algebra over $X$. The \emph{Eilenberg-Moore}
category $\cat^{\ir}$ for a monad $\ir$ consists of $\ir$-algebras and
their homomorphism. The \emph{Kleisli} category $\cat_{\ir}$ consists
of the same objects as $\cat$, but morphisms from $X$ to $Y$ in $\cat_{\ir}$ are
morphisms $X \to \uir Y$ in $\cat$. The Kleisli category $\cat_{\ir}$
inherits any coproducts $\cat$ has.
A strong monad
$\ir$ is \emph{commutative} when, for every
\[
     a : \ir X, b: \ir Y \tinf\ \
     \expdomon\ir {x \<- a;
      y \<- b;
      \!\sreturn \tpair xy}
\quad=\quad
     \expdomon\ir {      y \<- b;
      x \<- a;
      \!\sreturn \tpair xy}
\]
(The notion of strong/commutative monad is due to \cite{kock:strong-functors-and-monoidal-monads};
our formulation of algebras also appears in
\cite{marmolejo-wood:monads-as-extension-systems}.)
\paragraph{Biproducts.}
A zero object $\zero$ is both initial and terminal.
A category has \emph{(finite, countable, etc.)
  biproducts} if it has a zero object (and hence zero morphisms
$\zeromor_{X,Y}:X{\to}\zero{\to} Y$)
and the following canonical
morphisms are invertible:
\[
\textstyle  \coseq[i \in I]{\seq[j \in I]{\delta_{i,j}} }: \sum_{i \in
      I}X_i \to \prod_{j \in I} X_j
\qquad
\text{where: }
\delta_{i,i} \definedby
  \id[X_i], \
\delta_{i,j} \definedby
  \zeromor_{X_i, X_j}  \text{ for $i \neq j$}.
\]

\paragraph{Algebraic structure.}
Recall the notion of a \emph{commutative monoid} $\triple{M}1\cdot$ in
a category with finite products. We extend it to countably many
arguments. Let $\cat$ be a category with countable products. A
\emph{$\sigma$-monoid}
(see also \cite{haghverdi-scott:cat-GoI}) is a triple $\triple M0\Sigma$
consisting of: an object $M$; a morphism $0 : \terminal \to M$; and
a morphism $\Sigma : M^{\naturals} \to M$
such that:
\begin{itemize}
\item setting $\delta_0\! \definedby \id[M] : M \to M$ and
  $\delta_i \definedby 0 \compose !{} : M \to \terminal \to M$, $i > 0$,
  we have $\Sigma \compose \seq[i \in \naturals]{\delta_i} = \delta_0$; and
\item for every bijection
  $\phi : \naturals \isomorphic \naturals\times\naturals$,
\(
    a_{\tpair - -} : M^{\naturals \times \naturals} \tinf
    \Sigma\seq[i \in \naturals]{\Sigma\seq[j \in \naturals]{a_{\tpair i j}}}
      = \Sigma\seq[k \in \naturals]{a_{\phi(k)}}
\text.
\)
\end{itemize}

\begin{proposition}
  In a category with countable biproducts, each object
  $M$ is a $\sigma$-monoid via:
  \begin{mathpar}
  \textstyle  \zeromor_{\terminal, M} : \terminal \to M

    \Sigma : \prod_{i \in \naturals}M \isomorphic \sum_{i \in
      \naturals}M \xto{\codiag} M
\text{ where $\codiag$ is the codiagonal.}
  \end{mathpar}
  Every morphism is a $\sigma$-monoid homomorphism with respect to
  this structure.
\end{proposition}
A \emph{$\sigma$-semiring} is a quintuple
$\seq{S, 1, \cdot, 0, \Sigma}$ consisting of:
a commutative monoid $\triple S1\cdot$; and
a $\sigma$-monoid $\triple S0\Sigma$, such that
\(
a : S, b_{-} \in S^{\naturals} \tinf
a\cdot \Sigma\seq[i \in \naturals]{b_i}
=
\Sigma\seq[i \in \naturals]{a\cdot b_i}
\).
Given a $\sigma$-semiring $\seq{S, 1, \cdot, 0, \Sigma}$, an
\emph{$S$-module} is a pair $\pair M\odot$ consisting of a
$\sigma$-monoid $M$; and a morphism $\odot : S\times M \to M$ satisfying:
\(x = x, 0_S \odot x = 0_M,
(a\cdot b)\odot x = a\odot (b \odot x),
   \parent{\Sigma\seq[n \in \naturals]{a_n}}\odot x
   =
   \Sigma\seq[n \in \naturals]{a_n \odot x}
\).
 \subsection{Synthetic measure theory}
Synthetic mathematics identifies structure and axioms from which we can
recover the main concepts and results of specific mathematical
theories, and transport them to new settings. We now briefly recount
the relevant parts of
\citepos{kock:commutative-monads-as-a-theory-of-distributions}{
  development}.
(In the finite discrete case, this is also related to
\citepos{jacobs-effectus}{ work on effectuses}.)

\subsubsection{Axioms and structure}

Let $\cat$ be a cartesian closed category with countable products and
coproducts, and let $\monad{\Mea}$ be a commutative monad over
$\cat$. If the morphism $! : \Mea\initial \to \terminal$ is invertible,
then both the Eilenberg-Moore category $\cat^{\monad\Mea}$ and the
Kleisli category $\cat_{\monad\Mea}$ have zero objects. As a
consequence, we have a canonical $\monad\Mea$-homomorphism
\(
\bind\coseq[i]{\seq[j]{\delta_{i,j}}} : \Mea \sum_{i \in
  \naturals}X_i
\to \prod_{j \in \naturals} \Mea X_j
\).

\begin{definition}
  A \emph{measure category} is a pair $\pair\cat{\monad\Mea}$
  consisting of a cartesian closed category $\cat$ with countable
  products and coproducts, and equalisers; and a commutative monad
  $\monad\Mea$ over $\cat$ such that
  the morphisms
  $! : \Mea\initial \to \terminal$
  and
  $\bind\coseq[i]{\seq[j]{\delta_{i, j}}} : \Mea\sum_{i \in \naturals}X_i
  \to \prod_{j \in \naturals}\Mea X_j$
  are invertible.

\end{definition}

We fix a measure category $\pair\cat{\monad\Mea}$ for the remainder of
this section. The intuition is that $\Mea X$ is the object of
distributions/measures over $X$. Kock shows that, while short, the
above definition has surprisingly many consequences.

Both the Eilenberg-Moore and the Kleisli categories have countable
biproducts, and as a consequence, all $\monad\Mea$-algebras have a
$\sigma$-monoid structure and all $\monad\Mea$-homomorphisms are
$\sigma$-monoid homomorphisms with respect to it. Moreover, this
structure on the free algebra on the terminal object $R \definedby
\Mea\terminal$ extends to a $\sigma$-semiring structure by setting: $
1 \definedby \sreturn \tUnit$ and $r\cdot s \definedby
\expdomon{\monad\Mea}{r; s}$.
Kock calls this structure the \emph{$\sigma$-semiring of scalars}.
Each $\monad\Mea$-algebra $\alg$ has an $R$-module structure:
\[
  r : R, a : \carrier\alg \tinf r \odot a \definedby \expdomon{\monad\Mea}{r; a}
\]
As $\cat$ has equalisers, for each object $X$, we may form the
equaliser
 \( \Prob X \xdasharrow{\sub_X} \Mea X \xtoto[\constantly 1]{\Mea !} R\)
 because $R = \Mea\terminal$.
Each $\sub_X$ is monic, the monadic structure factors through $\sub$
turning $\Prob$ into a commutative monad $\monad\Prob$, and
$\sub : \monad\Prob \monomo \monad\Mea$ into a strong monad monomorphism.

The morphism $\Mea ! : \Mea X \to R$ is called the \emph{total measure}
morphism, and $\Prob$ is then the sub-object of all the measures with
total measure $1$, and so we think of it as the object of
\emph{probability measures} over $X$.  For example, every
$\monad\Prob$-algebra is closed under \emph{convex} linear
combinations of scalars: if $r_{-} : \naturals \to R$ satisfies
$\Sigma \seq[i] {r_i} = 1$ then
$\qmu_- : (\Prob X)^{\naturals} \tinf \Mea !(\Sigma\seq[i]{r_i\odot \qmu_i})
= 1$.

\subsubsection{Notation and basic properties}

\begin{figure}
  \begin{notation}
  \noassumptions       R & \Mea \terminal
  & \text{Scalars}
  \\
  f : Y^X, \qmu : \Mea X
  & f_*\qmu
  & (\Mea f)(\qmu)
  & \text{Push-forward}
  \\
  \qmu : \Mea X
  & \qmu(X)
  & !_*\qmu & \text{The total measure}
  \\
  x : X
  & \qdelta_x
  & \sreturn(x)
  & \text{Dirac distribution}
  \\
  \qmu : \Mea X, f : (\Mea Y)^X
  & \qint_X f(x)\qmu(\dif x)
  & \qmu \bind f
  & \text{Kock integral}
  \\
  w : R^X, \qmu : \Mea X
  &  w \odot \qmu
  & \qint_X (w(x)\odot \qdelta_x)\qmu(\dif x)
  & \text{Rescaling}
  \\
  \bracks{
  \begin{aligned}
      &f : (TZ)^{X\times Y},
    \\& x : X, k : (TY)^X
  \end{aligned}}
  &\qint_{Y} f(x,y)k(x, \dif y)
  &\qint_Y f(x,y)k(x)(\dif y)
  & \text{Kernel integration}
  \\[10pt]
  \bracks{
  \begin{aligned}
      &f : (\Mea X)^{X\times Y },
    \\&
    \qmu \in \Mea (X\times Y)
  \end{aligned}}
  & \qiint_{X\times Y} f(x,y)\qmu(\dif x,\dif y)
  & \qint_{X\times Y} f(z)\qmu(\dif z)
  & \text{Iterated integrals}
  \\
  \qmu : \Mea X, \qmu[2] : \Mea Y
  &  \qmu\tensor\qmu[2]
  & \qint_X\parent{\qint_Y\qdelta_{\tpair xy}\qmu[2](\dif y)}\qmu(\dif x)
  & \text{Product measure}
  \\
    \qmu : \Mea X, f : \carrier \alg^X
  & \Exp[\alg]_{x\sampled\qmu}[f(x)]
  & \qmu \bind f
  & \text{Expectation}
  \\
  f : R^X, \qmu : \Mea X
  & \int_X f(x)\qmu(\dif x)
  & \Exp[R]_{x \sampled \qmu}[f(x)]
  & \text{Lebesgue integral}
\end{notation}
 \caption{Synthetic measure theory notation}\figlabel{notation}
\end{figure}

Kock's theory shines brightly when we adopt a measure-theoretic
notation, as in \figref{notation}, by thinking of $\Mea X$ as the
object of measures over $X$, and $R$ as the object of scalars these
measures take values in. The functorial action of the monad  allows us to
push measures along morphisms, and pushing all the measure into the
terminal object gives a scalar we think of as the total measure of an
object. The monadic $\return$ acts as a dirac distribution. The main
advantage is the \emph{Kock integral}, synonymous to the monadic
$\bind$. The main difference between the Kock integral $\qint$ and the
usual Lebesgue integral $\int$ from measure theory is that the
Kock integral evaluates to a \emph{measure}, and not a scalar. Calculating with
the Kock integral is analogous to using Lebesgue integrals with
respect to a generic test function, and proceeding by algebraic
manipulation. The scalar rescaling $\odot$ allows us to rescale a
distribution by an arbitrary weight function. A \emph{kernel} is a
morphism $k : X \to \Mea Y$, and we use the usual notation for integration
against a kernel and iterated integration. We define the product
measure by iterated integration. Finally, the $\bind$ operation of an
$\monad\Mea$-algebra $\alg$ gives rise to an expectation operation.
Here we will only make use of the scalars' algebra structure, which
generalises the usual Lebesgue integral.

The justification for this notation is that it obeys the expected
properties, which we now survey. The commutativity of the monad lets
us change the order of integration:
\begin{theorem}[Fubini-Tonelli]
  For every pair of objects $X$, $Y$ in a measure category
  $\tpair \cat{\monad\Mea}$:
  \begin{multline*}
\textstyle    \qiint_{X\times Y}f(x,y)(\qmu\tensor\qmu[2])(\dif x, \dif y)
    =
    \qint_{Y}\qmu[2](\dif y)\qint_{X}\qmu(\dif x)f(x,y)
    =
    \qint_{Y}\qmu[2](\dif y)\qint_{X}\qmu(\dif x)f(x,y)
  \end{multline*}
  Moreover, for every $\Mea$-algebra $\alg$:
  \[
    \qmu : \Mea X, \qmu[2] : \Mea Y, f : \carrier\alg^{X\times Y}
    \tinf
    \Exp[\alg]_{\substack{x\sampled\qmu\\y\sampled\qmu[2]}}\,[f(x,y)]
    = \Exp[\alg]_{x\sampled\qmu}\,[\Exp[\alg]_{y\sampled\qmu[2]}\,[f(x,y)]]
    = \Exp[\alg]_{y\sampled\qmu[2]}\,[\Exp[\alg]_{x\sampled\qmu}\,[f(x,y)]]
  \]
\end{theorem}
As usual, we allow placing the binder $\qmu(\dif x)$ on either side of
the integrand $f(x)$.

Integrals and expectation interact well with the $R$-module structure in the sense that they are homomorphisms in both arguments. The precise statement of this fact can be found in Appendix \ref{sec:kock-equations}.

The push-forward operation interacts with rescaling in the following
way:
\begin{theorem}[Frobenius reciprocity]
  For all objects $X$, $Y$ in a measure category $\pair \cat{\monad\Mea}$:
  \[
\textstyle    w : R^X, \qmu : \Mea X, f : Y^X \tinf w \odot \parent{f_*\qmu}
    =
    f_*\parent{(w \compose f)\odot\qmu}
  \]
\end{theorem}

When calculating in this notation, we use the equations in
Appendix \ref{sec:kock-equations} where we present a toolbox for synthetic measure theory. This toolbox includes most of the equations we come to expect from standard measure theory, like the change of variables law. To validate them, inline the definitions and
proceed using the usual category-theoretic properties.

The following two sections contain relevant extensions to Kock's
theory.

\subsubsection{Radon-Nikodym derivatives}\subsubseclabel{Radon-Nikodym}
The Radon-Nikodym Theorem is a powerful tool in measure theory, and we
now phrase a synthetic counterpart. As usual in the synthetic setting,
we set the definitions up such that the theorem will be true. Doing so
highlights the difference between three measure-theoretic concepts
that coincide in measure theory, but may differ in the synthetic setting.

Let $\qmu, \qmu[2] \in \Mea X$ be measures. We say that $\qmu[2]$ is
\emph{absolutely continuous} with respect to $\qmu$, and write
$\qmu[2] \absc \qmu$, when there exists a morphism $w : X \to R$ such
that $\qmu[2] = w\odot\qmu$. Given two morphisms $w, v : X \to R$ and
a measure $\qmu \in \Mea X$, we say that $w$ and $v$ are \emph{equal
  $\qmu$-almost everywhere} ($\qmu$-a.e.) when
$w \odot \qmu = v \odot \qmu$. A \emph{measurable property} over $X$
is a morphism $P : X \to \boolty$. Given a measure $\qmu \in \Mea X$ a
measurable property $P$ over $X$ \emph{holds $\qmu$-a.e.}, when the
morphism $[P] \definedby \fun x\tif{P\ x}{1}{0}$ is equal
$\qmu$-a.e. to $1$.

\begin{theorem}[Radon-Nikodym]
  Let $\pair \cat\Mea$ be a well-pointed measure category.  For every
  $\qmu[2] \absc \qmu$ in $\Mea X$, there exists a $\qmu$-a.e.~unique
  morphism $\der {\qmu[2]}{\qmu} : X \to R$ satisfying
  $\der {\qmu[2]}{\qmu} \odot \qmu = \qmu[2]$.
\end{theorem}

\subsubsection{Kernels}
We say that a kernel $k: X\to \Mea Y$ is \emph{Markov} when, for all $x$,
$k(x, Y) = 1$, i.e., when $k$ factors through the object of probability
measures via $\sub : \Prob \monomo \Mea$. We now restrict attention to
kernels $k : X \to \Mea X$ over the same object $X$.  We say that such
a kernel \emph{preserves} a measure $\qmu$ when $\qmu \bind k = \qmu$.
Recall the morphism
$\braid[] \definedby \fun{\tpair xy}{\tpair yx} : X\times Y \to Y\times
X$. Given a measure $\qmu \in \Mea X$ and a kernel $k$, we define the
\emph{box product} by
$\qmu \bprod k \definedby \qiint_{X\times X}\qdelta_{\pair
  xy}\qmu(\dif x)k(x, \dif y)$.  A kernel $k$ is \emph{reversible}
with respect to a measure $\qmu \in \Mea X$ when
$\braid[]_*(\qmu\bprod k) = \qmu\bprod k$.

The following standard results on kernels transfer into the synthetic
setting. If a \emph{Markov} kernel $k$ is reversible with respect to
$\qmu$, then $k$ preserves $\qmu$. Kernels obtained by rescaling the
Dirac kernel, i.e., $\fun x w(x) \odot \qdelta_x$ are reversible
w.r.t.~all measures. Finally, linear combinations
$\fun x\sum_{n \in \naturals}r_n\odot k_n(x)$ of reversible kernels
w.r.t.~$\qmu$ are also reversible w.r.t.~$\qmu$.
 \subsection{Quasi-Borel spaces}\subseclabel{qbs}
It remains to show that there is a concrete model of synthetic measure theory
that contains the classical measure theoretic ideas that are central to probability theory and inference.
This is novel because \citepos*{kock:commutative-monads-as-a-theory-of-distributions}{work} is targeted
at the geometric/topological setting, whereas probability theory is based around Borel sets rather than open sets.
It is non-trivial because the traditional setting for measure theory does not support higher-order functions~\cite{aumann:functionspaces} and
commutativity of integration is subtle in general.
In this section we resolve these problems by combining some
recent discoveries \cite{HeunenKSY-lics17,sfinite},
and exhibit a model of synthetic measure theory which contains classical measure theory, for instance:
\begin{itemize}
\item the $\sigma$-semiring over the morphisms $\terminal \to R$
  is isomorphic to the usual $\sigma$-semiring over the extended non-negative reals, $\ereals+$;
\item this isomorphism induces a bijective correspondence between the
  morphisms $R\to \terminal +\terminal $ and the Borel subsets of
  $\ereals+$, as characteristic functions, and also between the
  morphisms $R\to R$ and the measurable functions $\ereals+\to
  \ereals+$;
\item it also induces an injection of the morphisms $\terminal
  \to \Mea(R)$ into the set of Borel measures on $\ereals+$, whose
  image contains all the probability measures;
  the morphisms $R\to \Mea(R)$ include all the Borel probability
  kernels;
\item the canonical morphism $R^{R}\times \Mea(R)\to R$, $(f,\qmu)\mapsto \int f(x)\,\qmu(\dif x)$, corresponds to classical Lebesgue integration.
\end{itemize}
Moreover, each object $X$ can be seen as a
set $U(X)=\cat(\terminal ,X)$ with structure,
because the category is well-pointed, in the sense that
the morphisms $X\to Y$ are a subset of the functions $U(X)\to U(Y)$.

\subsubsection{Rudiments of classical measure theory}
Measurable spaces are the cornerstone of conventional measure theory, supporting a notion of measure.

Recall that a \emph{$\sigma$-algebra} on a set $X$ is a set $\Sigma_X$ of subsets of $X$ that is closed under countable unions and complements.
A \emph{measurable space} is a set together with a $\sigma$-algebra.
A \emph{measure} is a $\sigma$-additive function $\Sigma_X\to \ereals+$.
A function $f$ between measurable spaces is \emph{measurable} if the inverse image
of every measurable set according to $f$ is measurable.

For example, on a Euclidean space $\ereals^n$ we can consider the Borel sets, which form the smallest
$\sigma$-algebra containing the open cubes. There is a canonical measure on $\ereals^n$, the \emph{Lebesgue measure},
which  assigns to each cube its volume, and thus to every measurable function $f\colon \ereals^n\to \ereals+$
a Lebesgue integral $\int_{\ereals^n}f \in \ereals+$.
A slightly more general class of measures is the \emph{$\sigma$-finite measures}, which include the Lebesgue measures and
are closed under disjoint unions and product measures.

A measurable space that arises from the Borel sets of a Polish space is called a \emph{standard Borel space}.
In fact, every standard Borel space is either countable or isomorphic to $\ereals$.
Standard Borel spaces are closed under countable products and countable disjoint unions.

\subsubsection{Quasi-Borel spaces}
\newcommand{\baseR}{\mathfrak R}
\newcommand{\QBS}{\textbf{QBS}}
In this section we fix an uncountable standard Borel space, $\baseR$.
For example, $\baseR=\ereals$.
The basic idea of quasi-Borel spaces is that rather than focusing
on measurable sets of a set $X$, as in classical measure theory,
one should focus on the admissible random elements $\baseR\to X$.
\begin{definition}[\cite{HeunenKSY-lics17}]
A \emph{quasi-Borel space (QBS)} is a set $X$ together with
a set of functions $M_X\subseteq [\baseR,X]$
such that $(i)$~all the constant functions are in $M_X$, $(ii)$~$M_X$ is closed under
precomposition with measurable functions on $\baseR$, and $(iii)$~$M_X$ satisfies the piecewise condition:
if $\baseR=\biguplus_{i=1}^\infty U_i$, where $U_i$ is Borel measurable
and $\alpha_i\in M_X$ for all $i$, then $\biguplus_{i=1}^\infty \alpha_i\cap (U_i\times X)$ is in $M_X$.

A \emph{morphism} $f\colon X\to Y$ is a function that respects the structure,
i.e.~if $\alpha\in M_X$ then $(f \compose \alpha)\in M_Y$.
Morphisms compose as functions, and we have a category $\QBS$.

A QBS $X$ is a \emph{subspace} of a QBS $Y$ if $X\subseteq Y$ and $M_X=\set{\alpha:\baseR\to X~\suchthat~\alpha\in M_Y}$.
\end{definition}
A measurable space $X$ can be turned into a QBS when given
the set of measurable functions $\baseR \to X$ as $M_X$. When $X$ and $Y$ are standard Borel spaces
considered as QBSes this way, $\QBS(X,Y)$ comprises the measurable functions, so
$\QBS$ can be thought of as a conservative extension of the universe of standard Borel spaces.
The three conditions on quasi-Borel spaces ensure that coproducts and products of standard Borel spaces
retain their universal properties in $\QBS$.
In fact, the category of $\QBS$s has all limits and colimits.
It is also cartesian closed; e.g., $\reals^{\reals}\definedby\QBS(\reals,\reals)$,
and $M_{(\reals^{\reals})}= \set{\alpha:\baseR\to(\reals^{\reals})~\suchthat~\mathrm{uncurry}(\alpha)\in\QBS(\baseR\times\reals\to\reals)}$.
For any QBS $X$, $M_X=\QBS(\baseR,X)$.

\subsubsection{A monad of measures} The following development is novel.
\begin{definition}\label{def:measure}
A \emph{measure} $\mu$ on a quasi-Borel space is a triple $(\Omega,\alpha,\mu)$
where $\Omega$ is a standard Borel space, $\alpha\in\QBS(\Omega,X)$, and $\mu$ is a $\sigma$-finite measure on $\Omega$.
\end{definition}
For example, $\Omega$ might be $\ereals ^n$ and $\mu$ might be the Lebesgue measure.
A measure determines an integration operator: if $f\in\QBS(X,\ereals+)$ then define
\[\textstyle\int f \dif\,(\Omega,\alpha,\mu)\definedby \int_\Omega f(\alpha(x))\,\mu(\dif x)\]
using Lebesgue integration according to $\mu$.
We say that two measures are equal, denoted $(\Omega,\alpha,\mu)\approx(\Omega',\alpha',\mu')$, if they determine the same integration operator.
We write $[\Omega,\alpha,\mu]$ for an equivalence class of measures.

As an aside,  we note that not every integration operator on $\ereals$ in the classical sense is a measure in the sense of Def.~\ref{def:measure},
because we restrict to $\sigma$-finite $\mu$.
Technically, the only integration operators that arise in this way are those corresponding to s-finite measures. This is a class of measures that includes the probability measures, and which works well with iterated integration
and probabilistic programming~\cite{sfinite}.

The measures up-to $\approx$ form a monad, as follows.
First, the set of all measures $\Qmonad X$ forms a QBS by setting
$M_{\Qmonad X}=\set{\fun r [D_r,\alpha(r,-),\mu|_{D_r}]~\suchthat~
\mu\text{\,$\sigma$-finite on $\Omega$},\,D\subseteq \baseR\times \Omega\,\text{measurable},\,\alpha\in \QBS(D, X)}$,
where $D_r=\{\omega~|~(r,\omega)\in D\}$.
In consequence, when $\Omega'$ is a standard Borel space,
for every morphism $f:\Omega'\to \Qmonad X$,
there exist $\Omega$, $\mu$, $D\subseteq \Omega'\times \Omega$ and $\alpha\in \QBS(D, X)$
such that $f(\omega')=[D_{\omega'},\alpha(\omega',-),\mu|_{D_{\omega'}}]$.
One intuition is that $\alpha$ is a partial function $\Omega'\times \Omega\to X$, with domain $D$.

The unit of the monad, $\return:X\to \Qmonad X$, is $\return(x)\definedby[\terminal,\fun{\_} x,\delta_{\tUnit}]$
where $\delta_{\tUnit}$ is the Dirac measure on the one-point space $\terminal$.
We often write $\qdelta_x$ for $\return(x)$.
The bind $\bind:\Qmonad X\times \Qmonad Y^X\to \Qmonad Y$ is
\[
[\Omega,\alpha,\mu]\bind \phi\ \definedby \ [D,\beta,(\mu\otimes \mu')|_D]
\]
where $\phi(\alpha(r))=[D_r,\beta(r,-),\mu']$. Note that $(\phi \compose \alpha):\Omega\to \Qmonad X$ must be of this form because it is a morphism from a standard Borel space.
The measure $\mu\otimes \mu'$ is the product measure, which exists because $\mu$ and $\mu'$ are $\sigma$-finite.

This structure satisfies the monad laws, it is commutative by the
Fubini-Tonelli theorem, and it satisfies the biproduct axioms, and so
it is a model of synthetic measure theory.  Every measure on
$\terminal$ is equivalent to one of the form $([0,r],!,\mu)$ where
$r\in \ereals+$, $! : [0,r] \to \terminal$ is the unique such random
element, and $\mu$ is the Lebesgue measure.  Thus $\Qmonad \terminal
\cong \ereals+$.

As another aside, we note that when $\Omega,\Omega'$ are standard Borel spaces,
the Kleisli morphisms $\Omega\to \Qmonad \Omega'$ correspond to s-finite kernels,
which were shown in \cite{sfinite} to provide a fully complete model of first-order probabilistic programming.
  \section{Continuous inference}\seclabel{continuous}
We now develop the continuous counterpart to \secref{discrete}. The
semantic structure of the category of quasi-Borel spaces allows
us to transport many of the definitions with little change. For example, a
monadic interface $\ir$ consists of analogous data, but the
assignments are indexed by quasi-Borel spaces, $\uir$ assigns
quasi-Borel spaces, and $\return^{\ir}$ and $\bind^{\ir}$ assign
quasi-Borel space morphisms.

\begin{definition}
  A \emph{continuous representation} $\ir$ is a tuple
  \(
    \seq*{\uir, \return^{\ir}, \bind^{\ir}, \mean^{\ir}}
  \)
  consisting of:
  \begin{itemize}
  \item a monadic interface $\triple*{\uir}{\return^{\ir}}{\bind^{\ir}}$;
  \item an assignment of a \emph{meaning} morphism
    $\mean^{\ir}_X : \uir X \to \Mea X$ for every space $X$
  \end{itemize}
  such that $\mean^{\ir}$ preserves $\return^{\ir}$ and $\bind^{\ir}$.

  A \emph{sampling} representation is a tuple
  $\seq*{\uir, \return^{\ir}, \bind^{\ir}, \mean^{\ir},
    \sample^{\ir}}$ such that its first four components form a continuous
  representation, it has an additional $\Qbs$-morphism
  $\sample^{\ir} : \terminal \to \uir\I$, and
  $\mean^{\ir}$ maps $\sample^{\ir}\tUnit$ to
  the uniform $\Qbs$-measure $\Uniform[] = [\I,\id,\UniformD]$
  on the unit interval $\I$, where $\UniformD$ is the usual
  uniform distribution on $\I$.

  A \emph{conditioning} representation $\ir$ is
  similarly a tuple
  \( \seq*{\uir, \return^{\ir}, \bind^{\ir}, \score^{\ir},
    \mean^{\ir}} \), with a $\Qbs$-morphism
  $\score^{\ir} : \reals+ \to \uir\terminal$ such that for each $r$,
  $\mean^{\ir}$ maps $\score^{\ir}(r)$ to the $r$-rescaled unit $\Qbs$-measure
  $r \odot \qdelta_{\tUnit} = [\terminal,\fun{\_} \tUnit, r \cdot \delta_{\tUnit}]$.

  An \emph{inference} representation $\ir$ is a
  tuple
  $\seq*{\uir, \return^{\ir}, \bind^{\ir}, \sample^{\ir},
    \score^{\ir}, \mean^{\ir}}$ with the appropriate components
  forming both a sampling representation and a conditioning representation.
\end{definition}
This definition refines \defref{disc inf rep} with sampling and
scoring representations, allowing us to talk about inference
transformers that augment a representation of one kind into another.

\begin{example}[Continuous sampler]
  By analogy with \exampleref{disc weighted sampler}, we define in
  \figref{sam} a sampling representation using the type
  \( \Sam \tvar \definedby \Variant{\Inj{Return} \tvar \vor
    \Inj{Sample} (\I \to \Sam\tvar)} \).  Validating the preservation of sample
  and the monadic interface is straightforward. It also follows from
  more general principles: $\monad\Sam$ is the initial monad with an
  operation $\sample : \uir\I$.
\end{example}

\begin{figure}
  \begin{minipage}[b]{.5\linewidth}
    \[
  \SampleMonad{\Sam}{
    \sreturn \var\ \ &= \Inj{Return} \var
  }{
    a \bind f        &= \imatch a{\Inj{Return} \var & f(x) \\
                                  \Inj{Sample} k    \\
                                  \multicolumn{2}{@{\quad}l@{}}{
                                      \Inj{Sample} (\fun r k(r) \bind f)}}
  }{
    \ssample         &= \Inj{Sample}{\fun r(\Inj{Return} r)}
  }{
    \mean a          &= \imatch a{\Inj{Return} \var & \qdelta_x \\[.2ex]
                                  \Inj{Sample} k    & \qint_{\ \I}k(x)\Uniform(\dif x)}
  }
\]
     \subcaption{Continuous sampler representation}\figlabel{sam}
  \end{minipage}
  \begin{minipage}[b]{.5\linewidth}
      \[
  \RepTrans{Rep}{Cond}{\Writer}
         {
           \sreturn_{\Writer\ir} x &= \return^{\ir} \tpair 1x
         }{
           a \bind_{\Writer\ir} f  &=
           \begin{aligned}[t]
             \expdomon\ir{&\tpair rx \<- a;
                        \\&\tpair sy \<- f(x);
                        \\&\sreturn \tpair{r\cdot s}y}
           \end{aligned}
         }{
           (\stmap \tfm)_X &= \tfm_{\reals+\t* X}
         }{
           \slift_{\ir} a &= \expdomon\ir{x \<- a; \sreturn \tpair 1x}
         }{
           \mean_{\Writer\ir} a              &=
              \fun x\qint_{\ \reals+\times X}r\odot
                   \qdelta_x m^{\ir}(a)(\dif r, \dif x)
           \\&
           \sscore_{\Writer\ir} r            &= \return^{\ir} \tpair r\tUnit
         }
  \]
     \medskip
    \subcaption{Continuous weighting inference transformer}
    \figlabel{cont weighting}
  \end{minipage}
  \caption{Continuous representations and tranformers}\figlabel{cont
    weight sampler}
\end{figure}

We define inference transformations between any two representations as
in \defref{disc inf tfm}. We have four kinds of representations, and
when defining transformers we can augment a representation with
additional capabilities:

\begin{definition}
  Let $k_1$, $k_2$ be a pair of kinds of representation.
  A \emph{$k_1$ to $k_2$ transformer} $\IF$ is a tuple
  $\seq*{\uIF, \itmap^{\IF}, \lift^{\IF}}$ consisting of an assignments
  of:
  \begin{itemize}
  \item a $k_2$ representation $\uIF \ir$ to every
    $k_1$ representation $\ir$;
  \item an inference transformation
    $\itmap^{\IF} \tfm : \IF\ir \to \IF\ir[2]$ to every
    transformation $\tfm : \ir \to \ir[2]$; and
  \item an inference transformation $\lift_{\uir} : \ir \to \uIF \ir$ to
    every $k_1$ representation $\ir$.
  \end{itemize}
  When the two kinds $k_1$, $k_2$ differ, we say that that the
  transformer is \emph{augmenting}.
\end{definition}

When defining a $k_1$ to $k_2$ transformer, we adopt a Haskell-like
type-class constraint notation $k_1 \implies k_2$ used for example in
\figref{suspend}.

\begin{example}\examplelabel{cont weighting trans}
  By analogy with \exampleref{disc weighting trans}, \figref{cont
    weighting} presents the \emph{continuous weighting} transformer
  structure on \( \Writer \ir\, X \definedby T(\reals+ \t* X) \). It
  augments any representation transformer with conditioning
  capabilities. Each conditioning operation is deferred to the return
  value, and so we can view this transformer as freely adding a
  conditioning operation that commutes with all other operations.
  When the starting representation had conditioning capabilities, we
  have an inference transformation $\waggr : \Writer\ir \to \ir $,
  given by
  \( \waggr a \definedby \expdomon\ir{ \tpair rx \<- a; \score^{\ir}
    r; \sreturn x } \) which conditions based on the aggregated
  weight.

  Its validity follows from a straightforward calculation using the
  meaning preservation of $\ir$.
\end{example}

In the continuous case, the output of the final inference transformation
will always be $\Writer \Sampling X$ or a similar $\Pop \Sampling X$ described
in the next section. From this representation, we obtain the Monte Carlo
approximation to the posterior by using a random number generator to supply
the values required by $\Sampling\!\!$. Interpreting the program directly in
$\Writer \Sampling X$ and sampling from that would correspond to simple
importance sampling from the prior, which usually needs a very large number
of samples to give a good approximation to the posterior. Our goal in approximate
Bayesian inference is therefore to find another representation for the program
and a sequence of inference transformations that map it to $\Writer \Sampling X$.
While, in principle, this output represents the same posterior distribution,
hopefully it uses a representation that requires fewer samples to obtain a good
approximation than a direct interpretation in $\Writer \Sampling X$. We emphasise
that approximation is only done in this final sampling step, while all the
inference transformations that happen before it are exact.
 \section{Sequential Monte Carlo}\seclabel{SMC}
\emph{Sequential Monte Carlo (SMC)} is a family of inference algorithms
for approximating sequences of distributions. In SMC, every distribution is
approximated by a collection of weighted samples called a \emph{population},
with each population obtained from the previous one in the sequence by a
suitable transformation. In a sub-class of SMC known as \emph{particle filters},
each distribution in the sequence is generated by a known random process from
the previous distribution and we can apply this process to samples in the
previous population to obtain the current population. In particle filters
the samples in the population are called \emph{particles}.

A common problem with particle filters is that, after multiple steps,
a few particles have much larger weights than the remaining ones,
effectively reducing the sample size in the population well below the actual
number of particles, a phenomenon known as \emph{sample impoverishment}. To ameliorate
this problem, particle filters introduce additional \emph{resample} operations
after each step in the sequence, which constructs a new population by sampling
with replacement from the old one. The new population has uniform weights
across its particles. In the setting of probabilistic
programming, we use suspended computations as particles, and their
associated weight is their currently accumulated likelihood.

We show how to decompose a particle filter into a stack of two
transformers: a representation to conditioning transformer for
representing a population of particles, and a conditioning to conditioning
transformer that allows us to run a particle until its
next conditioning operation. We define each step of the SMC algorithm
as an inference transformation on this stack.  We can then apply this
stack of transformers to a sampling representation to obtain a correct
by construction variation of SMC. The algorithm we obtain is known as
the particle filter with multinomial resampling \citep{doucet-johansen:smc-tutorial11}
that uses the prior as the proposal distribution, but throughout
this paper we refer to it simply as SMC.

\subsection{The population transformer}
Given a representation $\ir$, we define a representation structure
over $\Pop \ir\, X \definedby \uir (\List (\reals+ \t* X))$. We further
deconstruct this representation transformer as the composition of two
transformers: the continuous weighting transformer $\Writer$ from
\exampleref{cont weighting trans}, and Haskell's notorious $\ListT$
transformer.

The negative reputation associated to the transformer
$\ListT \ir\, X \definedby \uir (\List X)$ stems from its failure to
validate the monad laws when $\ir$ is not
commutative.\footnotemark\ However, it is a perfectly valid
representation transformer, described in \figref{listT}, since we do not
require that representations satisfy monad laws.

\footnotetext{For a list transformer ``done right'', see
  \citepos*{jaskelioff:thesis}{thesis}, and its
  generalisations~\cite{pirogy:em-monoids, fiore-saville:list-objects}.}

To prove the meaning function preserves $\return$, simply
calculate. For $\bind$ preservation, show:
\[
  a_s : \List(\uir X) \tinf
  \mean^{\ListT \ir} (\sequence a_s) = \sum_{a \in a_s} \mean^{\ir}(a)
\]
and proceed via straightforward calculation using the linearity of the
Kock integral and the commutative ($\sigma$-)monoid structure on measures.

\begin{figure}
  \begin{minipage}[b]{.5\linewidth}
    \[
  \begin{array}{@{}l@{}}
    \text{Auxiliary functions:} \\
    \sequence : \List (TX) \to T(\List X) \\
    \sequence  \definedby \begin{array}[t]{@{}l@{}l@{}}
      \foldlist &(\return \lst-) \\
                &(\fun {\tpair ar}
                    \expdomon\ir{\begin{array}[t]{@{}l@{}l@{}}
                                   x   &\<- a; \\
                                   x_s &\<- r;\\
                                   \multicolumn{2}{@{}l@{}}{
                                   \sreturn (x :: x_s)}}
                                 \end{array}\\
                    \end{array}
    \\
    \concat : \List(\List X) \to \List X\\
    \concat \definedby \foldlist {\lst-} \++
    \\
    \sum_{x \in x_s}f(x) \definedby
      \foldlist 0
         \parent{\fun {\tpair xs}f(x)+s}\ x_s
    \\\\
    \RepTrans{Rep}{Rep}{\ListT}{
      \sreturn_{\ListT \ir} x &{}= \return^{\ir} \lst x
      }{
        a \bind_{\ListT \ir} f &{}=
        \expdomon\ir{\begin{array}[t]{@{}l@{}}
            x_s \<- a; \\
            \letin {b_{s}}
              {\tmap f\ x_s}          \\
            y_{ss} \<- \sequence{b_s}; \\
            \sreturn (\concat y_{ss})
        }
                     \end{array}
      }{
        \mean_{\ListT \ir}a &= \qint_{\List X}
                              \mean^{\ir}(a)(\dif x_s)
                              \sum_{x \in x_s}\qdelta_x
      }{
        \slift_{\ListT \ir} a &= \expdomon\ir{\begin{array}[t]{@{}l@{}}
                                x \<- a;
                                \sreturn{\lst{x}}}
                                              \end{array}
      }{
        (\stmap \tfm)_X &= \tfm_{\List X}
      }
  \end{array}
\]
     \subcaption{The list transformer}\figlabel{listT}
  \end{minipage}
  ~~
  \begin{minipage}[b]{.5\linewidth}
    \(
  \begin{array}{@{}l@{}}
    \RepTrans{Rep}{Cond}{\Pop}{
      \sreturn_{\Pop\ir}{} &= \sreturn_{(\Writer \compose \ListT) \ir}
    }{
      \bind_{\Pop\ir}    &= {\bind_{(\Writer \compose \ListT) \ir}}
    }{
      \slift_{\Pop\ir}   &= \slift_{\Writer (\ListT \ir)} \compose \slift_{\ListT \ir}
    }{
      \stmap_{\Pop\ir}   &= \stmap_{\Writer (\ListT \ir)} \compose \stmap_{\ListT\ir}
    }{
      \mean_{\Pop\ir}    &= \mean_{(\Writer\compose\ListT)\ir}\\&
                       &= \fun a
                         \sqint\limits_{\mathclap{\List(\reals+\times X)}}
                         \mean^{\uir}(a)(\dif x_s)
                         \sum_{\pair rx\in x_s}r\odot \qdelta_x
      \\[3ex]&
      \sscore_{\Pop\ir} &= \sscore_{(\Writer\compose\ListT) \ir}
    }
\end{array}
\)
\subcaption{The population transformer}\figlabel{population}
\[
    \begin{array}{@{}l@{}}
    \replicate : \naturals \t* X \to \List X \\
    \replicate \tpair nx \definedby \tfold\naturals
                            \pfun{
                             \Inj{Zero}     &\To \lst- \\\vor
                             \Inj{Succ} x_s &\To x::x_s
                            \}\ n\gobble}
    \\
    \spark : \naturals_+ \To \Pop \ir\, \Unit \\
    \spark \definedby \return^{\ir}
        \parent{\replicate \tpair {n}{\tpair {\tfrac 1{n}}\tUnit}}
    \\
    \spawn : \naturals_+ \t* \Pop \ir\, X \to \Pop \ir\, X \\
    \spawn \tpair na \definedby
      \expdomon{\Pop\ir}{
      \begin{array}[t]{@{}l@{}}
        \spark n;
        a
      }
      \end{array}
  \end{array}
\]
     \subcaption{Spawning new particles}
    \figlabel{spawning}
  \end{minipage}
  \caption{Representing populations}
\end{figure}

By composing the two representation transformers, we obtain the
representation to conditioning transformer $\Pop$, given explicitly in
\figref{population}.

\figref{spawning} presents a $\naturals_+$-indexed family of inference
transformations. Fix any $n \in \naturals$. The $\spark$ function
generates a population of particles with the unit value, and
the same weight $\tfrac 1n$.  Thus, $\spawn\pair na$ takes a
distribution $a$ over particle populations, sparks $n$ equally
weighted particles, and for each of them, samples a population based
on $a$. A straightforward calculation confirms that the meaning of
$\spark$ is $1$, and so $\spawn\pair n- : \Pop \ir \to \Pop \ir$ is an
inference transformation.  In the version of SMC we consider below, we
will only pass to $\spawn$ a distribution $a$ over uniformly-weighted
single-particle populations.

We use $\spawn$ to \emph{resample} a new population. Thinking
operationally, we have a population of weighted particles and we
obtain a new population by sampling with replacement from the
current one, where the probability of selecting a given particle is
proportional to its weight. Doing so is equivalent to simulating a
discrete weighted sample using a uniform one.
\begin{lemma}\lemmalabel{randomisation}
  There is a $\Qbs$-morphism
  $\dwrand : \List(\reals+\t* X)\t*\I \to \Variant
  {\Inj{Take}X \vor \Inj{Fail}}$ such that:
  \begin{itemize}
  \item For all $x_s$ for which $\sum_{\tpair r\_ \in x_s}r = 0$, we
    have $\dwrand(x_s, -)_*\Uniform = \qdelta_{\Inj[]{Fail}}$.
  \item For all $x_s$ for which $w := \sum_{\tpair r\_ \in x_s}r > 0$,
    we have
    $\dwrand(x_s, -)_*\Uniform = \sum_{\tpair rx \in x_s} \tfrac
    {r_i}w\odot \qdelta_{\Inj{Take}x}$.
  \end{itemize}
\end{lemma}
\figref{dwrand} presents one such morphism, though its precise
implementation does not matter to our development.  As a consequence,
for every sampling representation $\ir$ for which we have an element
$\fail : TX$ such that $\mean^{\ir}(\fail) = 0$, we can define a
discrete weighted sampler $\dwsampler^{\ir}(x_s) : \List(\ereals+\*X) \to TX$ in \figref{dwsampler} which will then
satisfy $\mean^{\ir}(\dwsampler^{\ir}(x_s)) = \sum_{\tpair rx \in x_s}r\odot \qdelta_x$.

\begin{figure}
  \begin{minipage}[b]{.5\linewidth}
  \[
  \begin{array}[t]{@{}l@{}}
  \dwrand(x_s, r) \definedby \\
    \quad\begin{array}[t]{@{}l@{}}
  \letin {w}{\sum_{\tpair r\_ \in x_s}r}\\
  \begin{array}{@{}l@{}l@{}}
    \tif{&w = 0\\}
      {{}&\Inj[]{Fail}\\}
      &\begin{array}[t]{@{}l@{}l@{}} \foldlist &{\tpair{w\cdot r}{\Inj[]{Fail}}}\\
                 &(\fun {\tpair{\tpair sx}{\tpair\fuel\result}} \\
                      &\begin{array}{@{}l@{}l@{}}
                        \tif{&0 \leq \fuel < s\\}
                          {&\tpair{-1}{\Inj{Take}x}\\}
                          {&\ \scomment{potential underflow}}\\&\tpair{\fuel - s}{\result})
                          \\x_s
                    \end{array}
       \end{array}
  \end{array}
         \end{array}
\end{array}
\]\\\bigskip
   \subcaption{A discrete weighted randomiser}
  \figlabel{dwrand}
  \end{minipage}
  ~~
  \begin{minipage}[b]{.5\linewidth}
    \[
  \begin{array}[t]{@{}l@{}}
    \dwsampler^{\ir}(x_s)
    \definedby\\
    \quad
    \begin{array}[t]{@{}l@{}l@{}}
      \expdomon\ir{&\score (\sum_{\tpair r\_ \in x_s}r); \\
                   &r \<- \ssample; \\
                   &\imatch{\dwrand\pair {x_s}r}
                           {\Inj[]{Fail} & \fail \\
                        \mathllap{\vor{}}\Inj{Take}x  & \sreturn x\}}\gobble}
    \end{array}
  \end{array}
\]
     \subcaption{A discrete weighted sampler}
    \figlabel{dwsampler}

    \[
  \begin{array}{@{}l@{}}
    \resample : \naturals_+ \t* \Pop\ir\, X \to \Pop\ir\, X \\
    \resample\tpair na \definedby \\
    \qquad
    \expdomon\ir{\begin{array}[t]{@{}l@{}}
    x_s \<- a;\\
    \spawn \tpair n {\dwsampler^{\Pop\ir} x_s}
    }
                 \end{array}
  \end{array}
\]
     \subcaption{Resampling}\figlabel{resampling}
  \end{minipage}
  \caption{The resampling transformation}
\end{figure}

The resampling step in \figref{resampling} operationally takes the
current population, creates a computation/thunk that samples a single
particle from this population, and then spawns $n$ new particles that
are initialised with this thunk. The morphism $\resample\pair n- : \Pop \ir \to \Pop \ir$ is an
inference transformation because, as we know,
$\spawn\pair n- $ is one and $\dwsampler^{\Pop\ir}: \Pop \ir \to \Pop \ir$ samples
a population consisting of just a single unit weight particle with
a probability proportional to its renormalised weight in the original population.

\subsection{The suspension transformer} \label{sec:sus}
The second transformer in the SMC algorithm allows us to suspend
computation after each conditioning. The suspension transformer equips
the standard \emph{resumption} monad transformer
$\Sus \ir\, X \definedby \uir \Variant { \Inj{Return} X \vor \Inj{Yield}
  (\Sus \ir\, X)}$, presented in \figref{suspend}, with inference transformations.

\begin{figure}
  \begin{minipage}[t]{.5\linewidth}
    \[
  \begin{array}{@{}l@{}}
    \RepTrans{Cond}{Cond}{\Sus}{
    \sreturn_{\Sus\ir}x &{}= \return_{\ir} (\Inj{Return}x)
    }{
    a \bind_{\Sus\ir} f &{}=\begin{array}[t]{@{}l@{}l@{}}
                          &\fold {}(
                                   \fun b
                            \expdomon{\ir}{\\
                            &t \<- b;\\
                            &\imatch t{
                            \Inj{Return}x & f(x)\\
                            \mathllap{\vor{}}\Inj{Yield}c  & \Inj{Yield}c
                          \})}
                          \gobble}
                          \\&a
                            \end{array}
    }{
      \slift_{\Sus\ir} a &= \expdomon\ir{x \<- a; \return_{\Sus\ir}x}
    }{
      (\stmap_{\Sus\ir} \tfm)_X&= \tfold{\Sus\ir X}(\fun b \mean_{\ir[2]}(b))
    }{
      \mean_{\Sus\ir} a &= m_{\ir}(\finish_{\Sus\ir}(a))

      \\
      &\sscore r      &= \return_{\ir}({\Inj{Yield}\lift_{\Sus\ir}(\sscore r)})
    }
  \end{array}
\]
   \subcaption{The suspension transformer}
  \figlabel{suspend}
  \end{minipage}
  ~\quad
  \begin{minipage}[t]{.5\linewidth}
    \[
  \begin{array}{@{}l@{}l@{}}
    \adva_{\ir} &: \Sus \ir X \to \Sus \ir X \\
    \adva_{\ir} a &{}=
                        \expdomon\ir{\\
                        &\qquad t \<- a;\\
                        &\qquad\imatch t {
                        \Inj{Return}x & \return_{\ir}x \\
       \mathllap{\vor{}}\Inj{Yield} t & t
                        \}}\gobble}
    \\
    \finish_{\ir} &: \Sus \ir X \to \ir X \\
    \finish_{\ir}a &{}=
                         \fold \fun b \expdomon\ir{\\
                           &\qquad t \<- b;\\
                         &\qquad \imatch t{
                           \Inj{Return}x & \return x \\
        \mathllap{\vor{}}  \Inj{Yield} b & b\}
                         }\gobble
                         }
  \end{array}
\]
     \subcaption{Suspension operations}\figlabel{suspension ops}
  \end{minipage}
  \caption{The suspension transformation}
\end{figure}

The two transformations on suspended computations in
\figref{suspension ops} take one step, and complete the computation,
accordingly. As the meaning function for the transformed representation
returns the meaning the computation would have if it was allowed to
run to completion, these two operations do not change the meaning and
so form inference transformations.

We can now put all the components together:
\begin{theorem}
Let $\ir$ be a sampling representation. For every pair of natural
numbers $n$, $k$, the following composite forms an inference transformation:
\begin{multline*}
\smc^{\ir}_{n, k} \definedby
  (\Sus\compose \Pop) \ir
  \xto{\stmap_{\Sus}\spawn\tpair n-}
         (\Sus\compose \Pop) \ir
\\  \xto{(\adva \compose \stmap_{\Sus}\resample(n, -))^{\compose k}}
         (\Sus\compose \Pop) \ir
  \xto{\finish}
         \Pop \ir
\end{multline*}
\end{theorem}
In the above $(-)^{\compose -} :  X^X\times \naturals \to X^X$
denotes $n$-fold composition.
The transformation $\smc^{\ir}_{n,k}$ amounts to running the SMC
algorithm with $n$ particles for $k$ steps.
If the representation
$\ir$ is operational in nature, such as the continuous sampler $\Sam$,
we get a sequence of weighted values over the return type when we run
the resulting representation. By construction, the distribution on the
results, rescaled according to their final weights, would be identical
to the desired posterior distribution.

When the representation $\ir$ is not a commutative monad, like the
continuous sampler $\Sam$, the resulting representation $\Pop \ir$ is
not a monad: the monad laws do not hold. Therefore, to encompass
representations of $\Pop \ir$ one must generalise beyond monads.

\section{Trace Markov Chain Monte Carlo}\seclabel{MHG}
Markov Chain Monte Carlo (MCMC) algorithms operate by repeatedly using
a transition kernel to generate a new sample from a current one.
Thus they can be thought of as performing a
random walk around the space they are exploring. If the transition
kernel is well-behaved, they are guaranteed to preserve the
distribution.
A popular MCMC algorithm used for Bayesian inference is Metropolis-Hastings (MH),
where the transition kernel consists of a proposal kernel followed by a decision
to either accept the proposed sample or keep the old one.
The accept or reject step is used to correct for bias introduced by the
proposal kernel, thus producing a valid MCMC algorithm for a rich family
of proposal kernels.

MH is a general inference method, but it requires
specialised knowledge about the space on which they operate on. In the
context of a probabilistic programming language, the \emph{Trace MH}
algorithm replaces the unknown target space with the space of program
\emph{traces}, which are shared by all probabilistic programs. Thus,
Trace MH allows probabilistic programming language designers to
devise general-purpose kernels to effectively explore traces.

We analyse the the Trace MH as follows. First, we prove a
quasi-Borel space counterpart of the \emph{Metropolis-Hastings-Green
  (MHG) Theorem}, that forms the theoretical foundation for the
correctness of MH. We then present the \emph{tracing} representation
and show its validity. We present the Trace MH algorithm,
parameterised by a proposal kernel for traces, and give sufficient
conditions on this kernel for the resulting transformation to be
valid. We then give a concrete proposal kernel and show that it
satisfies these general conditions.

\subsection{Abstract Metropolis-Hastings-Green}
In the abstract, the key ingredient in MH is the
\emph{Metropolis-Hastings-Green (MHG)} morphism $\mh[]$ presented in
\figref{mh update}, formulated in terms of an arbitrary inference
representation $\ir$. This transformation is usually known as the
\emph{update step} of the MH algorithm.  It is parameterised by a
(representation of a) \emph{proposal} kernel $\psi : X \to \uir X$, and by
a chosen (representation of a) Radon-Nikodym derivative
$\rho : X\times X \to \ereals+$.

\begin{figure}
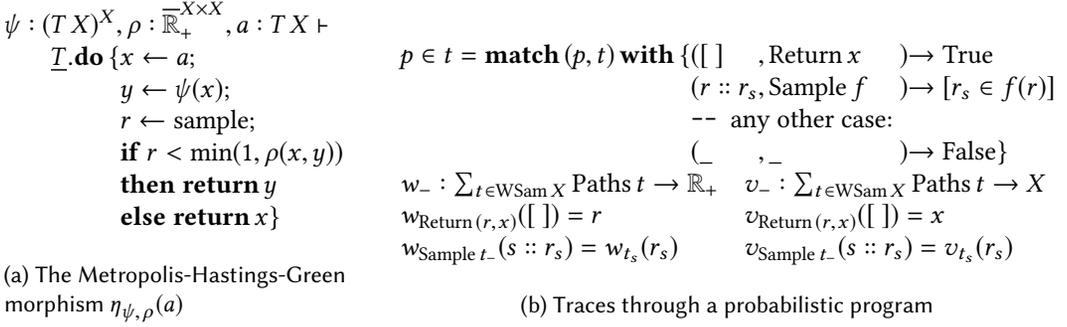

  \begin{minipage}[b]{.375\linewidth}
    \[
  \begin{array}[t]{@{}l@{}l@{\qquad}}
    \psi : (&\uir X)^{X},  {}\rho : \ereals+^{X\times X}, a : \uir X \tinf\\
    &\expdomon\ir{
                \begin{array}[t]{@{}l@{\,}l@{}}
                        x \<- a;  \\
                        y \<- \psi(x); \\
                        r \<- \ssample; \\
                        \tif {r < \min(1,\rho(x,y))\\}{
                        \sreturn y      \\}{
                        \sreturn x}}
                \end{array}
\end{array}
\]
     \subcaption{The Metropolis-Hastings-Green\\morphism $\mh[{\psi, \rho}](a)$}\figlabel{mh update}
  \end{minipage}
  ~
  \begin{minipage}[b]{.625\linewidth}
  \[
\begin{array}{@{}l@{}l@{}}
p \in t {}= \vmatch{\tpair pt}{
    \begin{array}[t]{*4{@{}l}@{\quad}l@{}}
	    \tpair{\lst{\ }&}{\Inj{Return}x&} &\To \tru \\
	    \tpair{r :: r_s&}{\Inj{Sample}f&} &\To [r_s \in f(r)] \\
            \multicolumn{2}{@{}c@{}}{\scomment{ any other case: }}\\
      \tpair{\_      &}{\_       &} &\To \fls} &
    \end{array}
  \\
  \begin{array}{@{}l@{\quad}l@{}}
    \weight \placeholder :  \sum_{t \in \WSam X} \paths t \to \reals+
    &
             \valuate\placeholder :  \sum_{t \in \WSam X} \paths t \to X\\
        \weight {\Inj{Return} \tpair rx}(\lst{\ }) = r &
                \valuate{\Inj{Return} \tpair rx}(\lst{\ }) = x\\
        \weight {\Inj{Sample}{t_{\placeholder}}}(s :: r_s) = \weight {t_s}(r_s)
        &
        \valuate{\Inj{Sample}{t_{\placeholder}}}(s :: r_s) = \valuate{t_s}(r_s)
  \end{array}
  \end{array}
\]
   \subcaption{Traces through a probabilistic program}\figlabel{traces}
  \end{minipage}
  \caption{Basic notions in Trace MH}
\end{figure}

To use $\mh[]$ in an inference transformation, we need to provide
well-behaved parameters $\psi, \rho$, and their behaviour may depend
on the representation of the input distribution $a$. In particular,
the parameter $\rho$ should represent a well-behaved appropriate
Radon-Nikodym derivative. To simplify our proofs, we also require that
the proposal kernel $\psi$ is Markov, which suffices for our
application.

\begin{theorem}[Metropolis-Hastings-Green]\theoremlabel{MHG update}
  Let $X$ be a qbs, $a \in TX$ a distribution, $\psi : X \to TX$ a
  kernel, and $\rho : X\times X \to \ereals+$ a $\Qbs$-morphism.  Set
  $k \definedby m^{\ir} \compose \psi$ and
  $\qmu \definedby [\rho\neq 0] \odot (m^{\ir}(a) \bprod k)$.

  Assume that:
        \begin{enumerate*}
                \item $k$ is Markov;
                \item $[1 = \parent{\rho \circ \braid[]} \cdot \rho]$ holds
                        $\qmu$-a.e.;
                \item $\rho$ is a Radon-Nikodym derivative of
                        $\braid[]_*\qmu$
                        with respect to $\qmu$; and
                \item $\rho(x,y) = 0 \iff \rho(y,x) = 0$ for all $x,y \in X$.
        \end{enumerate*}

        Then $(m^{\ir} \compose \mh[\psi,\rho])(a) = m^{\ir}(a)$.
\end{theorem}

Using Kock's synthetic measure theory, we were able to follow closely
standard measure-theoretic proofs of MHG \cite{geyer:intro-mcmc}.  The
synthetic setting highlights the different roles each of the three
abstractions: a.e.-equality, a.e.-properties, and Radon-Nykodim
derivatives play in the proof that our formulation exposes
(cf.~\subsubsecref{Radon-Nikodym}).

\subsection{Tracing representation}
A sampling \emph{trace} is a sequence of samples that occur during the
execution of a probabilistic program. We represent such programs as
elements of the continuous weighted sampler $\Writer \Sam$ from
(cf.~\figref{cont weight sampler}). Consequently, the collection of
traces through a program $t \in \Writer \Sam X$ is a subset of
$\List\I$. \figref{traces} defines a measurable predicate
$[\in] : \Writer \Sam X \times \List \I \to \boolty$ that tests whether
a given sequence $p$ of probabilistic choice forms a complete trace in
the program $t$. Consequently, we can define the set of \emph{paths}
through a given program $t$ by
$\paths t \definedby \set{p \in \List \I \suchthat p \in t} \subset
\List\I$, and equip it with the subspace structure it inherits from
$\List \I$. We can therefore define the set: \[
  \sum_{t \in \WSam X}\paths t := \set{\tpair tp \in \WSam X \times
    \List \I \suchthat p \in t} \subset \WSam X \times \List \I
\]
which we can also equip with a subspace structure. We can now define
the \emph{weight} $\weight-$ and \emph{valuation} $\valuate-$
morphisms in \figref{traces} that retrieve the likelihood and value at
the end of a trace.

We can now define the tracing inference representation. It is
parameterised by an inference representation $\ir$ and given for $X$
as the following subspace of $\WSam X \times T(\List \I)$:
\[
\Trace \ir\ X  \definedby
        \set{ \tpair t a \in \WSam X \times \uir(\List \I)
                      \suchthat
                      \begin{aligned}[<{\textstyle}c]
                        &\text{$[{}\in t]$ $\mean_{\ir}(a)$-a.e., and}
                        \\&\textstyle
                        m_{\WSam}(t)
                        =
                        \qint_{\List\I}
                          \qdelta_{\valuate{t}(p)} \mean_{\ir}(a)(\dif p)
                      \end{aligned}
                      }.
\]
Thus, a representation consists of a program representation $t$,
together with a distribution $a$ on all lists, but maintaining two
invariants. First, the lists are $\mean_{\ir}(a)$-almost-everywhere paths
through $t$, and so we can indeed think of $a$ as a representation
of a distribution over traces. Second, if we calculate the posterior
of the paths through $t$ according to $\mean_{\ir}(a)$, it should have the
same meaning as the original program.

We stress that an implementation need \emph{not} compute the meaning
of the program. But this representation guarantees that the meaning
will be preserved by the inference operations.

Note that the integrand in the definition of
$\tpair ta \in \Trace T\ X$ is only partially defined. This partiality
is not an issue because the first condition guarantees it is
$m^T(a)$-a.e.~defined. We can then choose the constantly $0$
distribution when $p \notin t$.

\begin{figure}
  \begin{minipage}[b]{.5\linewidth}
    \[
  \GenMonad{Inf \implies Inf\ Monad}{}{\Trace\ir}{
    &\sreturn x       &{}= \tpair{\return_{\WSam}x}{\return_{\ir}\lst-}
    \\
    &\tpair ta\bind \tpair{f}{g} &{}= \pair*{t \bind_{\WSam}f}{
        \expdomon{\ir}{      \begin{array}[t]{@{}l@{}}
                     p \<- a; \\
                     q \<- g\compose \valuate t(p); \\
                     \sreturn (p \++ q))

      }}\end{array}
    \\
    &\mean\tpair ta &{}= \mean_{\WSam}(t)
                          =                         \qint_{\List\I}
                          \qdelta_{\valuate{t}(p)} \mean_{\ir}(a)(\dif p)

    \\
    &\stmap \tfm &{}= \id\times \tfm_{\List\I}
    \\
    &\ssample &{}= \tpair{\begin{array}[t]{@{}l@{}}
      \sample_{\WSam}}
    {\\\expdomon\ir{r \<- \ssample; \sreturn \lst r}}
                          \end{array}
    \\
    &\sscore r &{}=\tpair{\begin{array}[t]{@{}l@{}}
                            \sscore_{\WSam}}
                            {\\\expdomon\ir{\sscore r; \sreturn \lst-}}
                          \end{array}
  }
\]
     \subcaption{The tracing inference}\figlabel{tracing rep}
  \end{minipage}
  ~
  \begin{minipage}[b]{.5\linewidth}
  \[
\begin{array}{@{}l@{}}
  \mh[\psi, \rho]^{\Trace T} : \Trace T\, X \to \Trace T\, X
  \\
  \mh[\psi, \rho]^{\Trace T}\pair ta \definedby
                                       \pair t{\mh[\psi_t, \rho_t](a)}
\end{array}
\]
   \subcaption{Trace MH update-step}\figlabel{trace mh}
  \[
\begin{array}{@{}l@{}}
      \iprior_{\ir}  :  \WSam X \to \uir(\List(\I)) \\
      \iprior_{\ir}(t) \definedby \fold \\\pfun{
  &\Inj{Return}\tpair rx &\To \return_{\ir} \lst- \\
  \vor & \Inj{Sample}k     &\To\expdomon\ir{\\&&r \<- \sample_{\ir};\\
                                        &&k(r)}}
\end{array}
\]
   \subcaption{Prior representation}\figlabel{prior}
  \end{minipage}
  \caption{Building blocks of Trace MH}
\end{figure}
\figref{tracing rep} presents the inference representation structure
of $\Trace \ir$.  Most of the proof revolves around preseving the
invariant, i.e., that these definitions define set-theoretic
functions.

The inference transformation
$\marginal_{\ir} : \Trace \ir X \to \ir X$ \emph{marginalises} the
trace transformer once it is no longer useful. It first samples a path
and then uses it to run the program discarding the weight:
\( \marginal \ \tpair ta = \domon{ x \<- a; \sreturn \ \valuate{t} (x)
} \). Its correctness is precisely the invariant.

\subsection{Inference with MHG}
The transition from $\ir$ to $\Trace \ir$ still requires a proposal
kernel and a representation of the appropriate derivative, but these
can now be given in terms of concrete traces.

Given an inference representation $\ir$, a \emph{trace proposal
  kernel} is a transformation representing a kernel
\( \psi : \parent{\sum_{t \in \WSam X} \paths t} \to T(\List \I) \).
A \emph{trace derivative} is a transformation representing the
derivative
\( \rho : \parent{\sum_{t \in \WSam X} \paths t \times \paths t} \to
\ereals+ \).  Given a trace proposal kernel $\psi$ and a trace
derivative $\rho$, \figref{trace mh} presents the trace MHG update
transformation using the corresponding MHG update on $T(\List\I)$.

The Trace MH update step requires some assumptions to form an
inference transformation:
\begin{theorem}[Trace Metropolis-Hastings-Green]\theoremlabel{Correctness of Trace MH}
  Let $T$ be an inference representation, $\psi$ a trace proposal
  kernel, and $\rho$ a trace derivative. Assume that, for every
  $\pair ta \in \Trace T\, X$, letting
  $k \definedby m^T\compose \psi_t$ and
  $\qmu \definedby [\rho_t\neq 0] \odot (m^T(a)\bprod k)$:
        \begin{enumerate*}
                \item $k$ is Markov;
                \item $[1 = \rho_t\cdot (\rho_t\compose \braid[])]$ holds
                        $\qmu$-a.e.;
                \item $\rho_t$ is a Radon-Nikodym derivative
                        of $\braid[]_*\qmu$
                        with respect to $\qmu$; and
                \item $\rho_t(p,q)=0 \iff \rho_t(q,p) = 0$ for all $p,q \in \List(\I)$.
        \end{enumerate*}
        Then $\mh[\psi, \rho]^{\Trace T} : \Trace T \to \Trace T$ is a valid
        inference transformation.
\end{theorem}

We will now demonstrate such a simple and generic trace proposal kernel
and trace derivative that implement a MHG update step of a popular
lightweight Metropolis-Hastings algorithm in several probabilistic
programming language systems
\cite{goodman_uai_2008, HurNRS15, wood-aistats-2014, Goodman2014}.

For any inference representation $\ir$, \figref{prior} defines the
morphism $\iprior_{\ir}$ that maps a representation $t \in \WSam X$ to
its prior distribution on paths over $t$.  Let
$\UD(n) \in \Mea(\naturals)$ be the measure for the uniform discrete
distribution with support $\{0,1,\ldots,n\}$.  Intuitively, it assigns
a probability $\frac{1}{n+1}$ to every element in the support.  It be
easily defined from $\sample_{\Mea}$, which denotes the uniform
distribution on $\I$, as in \lemmaref{randomisation}.

We now define our concrete proposal $\psi_t$ and derivative, a.k.a.~ratio, $\rho_t$:
\[
\begin{array}{@{}l@{}}
        \psi_t  : \List(\I) \to T(\List(\I)) \\
        \psi_t(p)  \definedby
                \expdomon\ir{\begin{array}[t]{@{}l@{}l@{}}
                        i\<- \UD^{\ir}(|p|)
                        \\
                        q\<-\iprior^T(\isubterm(t,\itake(i,p)))
                        \\
                        \sreturn(\itake(i,p)+ q)}
                \end{array}
\end{array}
\mspace{150mu}
\begin{array}{@{}l@{}}
        \rho_t  :  \List(\I) \times \List(\I) \to \ereals+ \\
        \rho_t(p,q)  \definedby \frac{\weight{t}(q) \cdot (|p|+1)}{\weight{t}(p) \cdot (|q|+1)}
\end{array}
\]
where $\isubterm(t, x)$ selects a subterm of a given term by following
the list $x$ and $\itake(i, p)$ retrieves the $i$-th prefix of
$p$. This proposal and derivative/ratio satisfy the condition in the Trace MH.

Our approach lets us combine MH updates with other inference building
block. For example, recall the SMC algorithm from Section
\ref{sec:sus}. Each time it performs resampling, multiple particles
are given the same values, which results in inadequate coverage of the
space, a phenomenon known as \emph{degeneracy}. One way to ameliorate
this problem is to apply multiple MH transitions to each particle
after resampling in order to spread them across the space, resulting
in an algorithm known as resample-move SMC
\cite{doucet-johansen:smc-tutorial11}.

The implemnetation of resample-move SMC is very similar to that of SMC from
Section \ref{sec:sus}, except we introduce an additional layer $\Trace$ between
$\Sus$ and $\Pop$:
\begin{theorem}
Let $\ir$ be a sampling representation. For every pair of natural
numbers $n$, $k$, $\ell$ the following composite forms an inference trasnformation:
\begin{multline*}
  \rmsmc^{\ir}_{n,k,\ell}\definedby
  (\Sus\compose \Trace\compose \Pop) \ir
  \xto{\stmap_{\Sus}\stmap_{\Trace}\spawn\tpair n-}
         (\Sus\compose \Trace\compose \Pop) \ir \\
  \xto{(\adva \compose \ \stmap_{\Sus}\mh[]^{\compose \ell} \ \compose \stmap_{\Sus}\stmap_{\Trace}\resample(n,-))^{\compose k}}
         (\Sus\compose \Trace\compose \Pop) \ir
  \xto{\marginal \compose \finish}
         \Pop \ir
\end{multline*}
\end{theorem}
In the above we apply $\ell$ MH transitions after each resampling.
Our compositional correctness criterion corresponds
to a known result that resample-move SMC is an unbiased importance sampler.

\section{Related work and concluding remarks}\seclabel{conclusion}
The idea of developing a programming language for machine learning and
statistics is old, and was explored at least in the early
2000s~\cite{BUGS,Ramsey-popl02,Park-popl05} as an interesting yet
niche research topic. In the past five years, however, designing such
a language and building its runtime system has become an active
research area, and lead to practical programming languages and
libraries~\cite{Stan,wood-aistats-2014,Goodman2014,Mansinghka-venture14,goodman_uai_2008,InferNet,Hakaru,Tran2017,Tabular,Murray2013}
. Most of these research efforts have focussed on developing efficient
inference algorithms and
implementations~\cite{Wingate2013,Le2017,Tran2017,Kucukelbir2015}.
Only a smaller amount of work has been dedicated to justifying the
algorithms or other runtime systems of those
languages~\cite{HurNRS15,BorgstromLGS-corr15}. Our work contributes to
this less-explored line of research by providing novel denotational
techniques and tools for specifying and verifying key components of
inference algorithms, in particular, those for expressive higher-order
probabilistic programming languages. Such specifications can then be
combined to construct the correctness argument of a complex
inference algorithm, as we have shown in the paper.

The idea of constructing inference algorithms by composing
transformations of an intermediate representations is, to the best of
our knowledge, relatively recent.  In previous work with
Gordon~\cite{Scibior2015}, we manipulated a free monad representation
to obtain an implementation of SMC. However, we did not implement MH,
did not break down SMC further into resampling and suspension, and our
semantics was not compositional. \citet{Zinkov2016} directly
manipulate syntax trees of a small language Hakaru. Their semantics is
only first-order and they focus on local program transformations
corresponding to solving integrals analytically, which is orthogonal
to our global transformations relating to sampling algorithms.

Our approach does not yet deal with two important aspects of
inference. In practice, one wants \emph{convergence} guarantees for
the inference algorithm, estimating the results within an error margin
after a given number of inference steps. As any purely-measure
theoretic approach, ours does not express such properties. Additionally,
we can not express algorithms that rely on derivatives of the density function
for the program traces, such as Hamiltonian Monte Carlo or
\emph{variational} inference. Developing a theory of differentiation over
quasi-Borel spaces might enable us to express such algorithms.

Another interesting direction for future work
is to develop a denotational account of some probabilistic programming
languages that allow users to select or compose parts of inference
algorithms~\cite{Mansinghka-venture14,Tran2017}. The exposure of an
inference algorithm in such languages breaks the usual abstraction of
probabilistic programs as distributions, and causes difficulties of
applying existing semantic techniques to such programs. Our more
intensional semantics may be able to overcome these difficulties.
Finally, it would be interesting to consider indexed or effect-annotated
versions of inference representations, transformations and
transformers, where indices or annotations ensure that inference
components are applied to a certain type of programs. Such a refined
version of our results may lead to a way of selectively applying
Hamiltonian Monte Carlo or other algorithms that assume the presence
of differentiable densities and a fixed length of all paths through the program.

\bibliography{references}

\appendix
\pagebreak
\section{A Toolbox for Synthetic Measure Theory}
\seclabel{kock-equations}
In this section, we list some equations that allow us to work in synthetic measure theory much as if we would in classical measure theory. In fact, the validity of these equations lets us port many classical measure theory proofs to our setting. Figure \ref{fig:kock properties} lists some basic equations that hold in synthetic measure theory.

\begin{figure}
  \[
\begin{array}[t]{@{}*3{l@{}}}
\qmu : \Mea X, a : Y^X &{}\tinf \qint_X \qdelta_{a(x)}\qmu(\dif x) &{}= a_*\qmu
\\
x_0 : X, f : (\Mea\!Y)^X &{}\tinf \qint_X f(x)\qdelta_{x_0}(\dif x)&{}= f(x_0)
\\
x_0 : X, f : \carrier\alg^X&{}\tinf
    \Exp_{x \sampled \qdelta_{x_0}}[f(x)]&{}= f(x_0)
\\
r : \Mea \terminal, \qmu : \Mea X
   &{}\tinf
\qint_{\terminal} \qmu\, r(\dif \xi)&{}= r\odot \qmu
\end{array}
\mspace{120mu}
\begin{array}[t]{@{}*3{l@{}}}
f : Y^X &{}\tinf f_*\qdelta_x = \qdelta_{f(x)}
\\
\qmu : \Mea X &{}\tinf \qint_X \qdelta_x\qmu(\dif x) = \qmu
\\
\qmu : \Mea X &{}\tinf \qmu(X)=\qint_X 1\qmu(\dif x)
\end{array}
\]
\[
\begin{array}[t]{@{}*3{l@{}}}
\qmu : \Mea X, a : Y^X, f : (\Mea Z)^Y &{}\tinf \qint_X (f\compose a)(x)\qmu(\dif x) &{}= \qint_Y f(y)(a_*\qmu)(\dif y)
\\
x_0 : X, w : R^X, f : (TY)^X &{}\tinf
\qint_Xf(x)(w\odot
        \qdelta_{x_0})(\dif x) &{}= w(x_0)\odot f(x_0)
\\
\qmu : \Mea X, w : R^X, f : (TY)^X &{}\tinf
\qint_Xf(x)(w\odot \qmu)(\dif x) &{}= \qint_X w(x) \odot f(x)\qmu(\dif x)
\\
\qmu : \Mea X, f : (\Mea Y)^X
&{}\tinf
\Exp_{x \sampled \qmu}^{\pair{KY}{\bind}}[f(x)] &{}= \qint_X f(x)\qmu(\dif x)
\\
\qmu : \Mea X, a : Y^X, f : \carrier\alg^Y
&{}\tinf
      \Exp_{y \sampled a_*\qmu}[f(y)] &{}= \Exp_{x \sampled
        \qmu}[(f\compose a)(x)]
\end{array}
\]

\[
\begin{array}{@{}*3{l@{}}}
\qmu : \Mea (X_1\times X_2), f : (\Mea Y)^{X_i}
&{}\tinf
\qiint_{X_1\times X_2}f(x)\qmu(\dif x,\dif y)
&{}=
\qint_{X}f(x)((\projection_i)_*\qmu)(\dif x)
\\
\qmu : \Mea X, f : (\Mea Y)^X, g : (\Mea Z)^Y
&{}\tinf
\qint_X\qmu(\dif x)\qint_Y g(y)f(x)(\dif y)
&{}=
\qint_Y g(y)\left(\qint_X f(x)\qmu(\dif x)\right)(\dif y)
\\
\qmu : \Mea X,
\qmu[2] : \Mea Y,
g : (\Mea Z)^Y
&{}\tinf
\qint_{Y}(\qmu\tensor g(y))\qmu[2](\dif y)
&{}=
\qmu \tensor
        \qint_{Y}g(y)\qmu[2](\dif y)
\end{array}
\]

\[
\begin{array}[t]{@{}l@{}l@{}r@{}l@{}}
\multicolumn{4}{@{}l@{}}{
\begin{array}{@{}l@{\quad}c@{\quad}c@{}}
\fbox{$\ctx \definedby \qmu_1 : X_1, \qmu_2 : X_2$},{i=1,2:}
&\ctx {}\tinf
(\projection_i)_*(\qmu_1\tensor\qmu_2) {}= \qmu_{3-i}(X_{3-i})\odot \qmu_i
&
\qint_X \qmu_1\ \qmu_2(\dif x) = \qmu_2(X_2) \odot \qmu_1
\end{array}
}\\[1ex]
\ctx, f : (\Mea Y)^{X_i} {}&{}\tinf{}&
\qiint_{X_1\times X_2}f(x_i)(\qmu_1\tensor\qmu_2)(\dif x_1,\dif x_2) &{}= {\qmu_{3-i}(X_{3-i})\odot\qint_{X_i}f(x_i)\qmu_i(\dif x_i)}
\\
\ctx, f^1 : Y_1^{X_1}, f^2 : Y_2^{X_2}
&{}\tinf{}&
(f^1\times f^2)_*(\qmu_1\tensor\qmu_2) &{}= (f^1_*\qmu_1)\tensor(f^2_*\qmu_2)
\\
\multicolumn{4}{@{}c@{}}{
  \qint_{X_1\times X_2} \parent{f^1(x_1)\tensor f^2(x_2)} (\qmu_1\!\tensor\qmu_2)(\dif x_1,\dif x_2)
  =
         \parent{\qint_{X_1} f^1(x_1) \qmu_1(\dif x_1)}\tensor
         \parent{\qint_{X_2} f^2(x_2) \qmu_2(\dif x_2)}}
\end{array}
\]
   \caption{Toolbox for synthetic measure theory}
  \figlabel{kock properties}
\end{figure}

Additionally, we have the following theorem which makes precise the statement that integration and expectation act as $R$-module homomorphisms.

\begin{theorem}[$\sigma$-linearity]\theoremlabel{sigma linearity}
  For all objects $X$, $Y$ in a measure category
  $\pair \cat{\monad\Mea}$ and $\monad\Mea$-algebras $\alg$:
  \[
    \begin{array}{@{}*4{>{\displaystyle}l@{}}}
    \qmu : \Mea X, r_{-} : R^{\naturals}, f_{-} : (\Mea Y)^{\naturals \times X}
    &{}\tinf{}&
    \qint_X\qmu(\dif x)\sum_{n \in \naturals}r_n\odot f_n(x)
      &{}=
      \sum_{n \in \naturals}r_n\odot \qint_X\qmu(\dif x)f_n(x)
    \\
      \qmu_- : (\Mea X)^{\naturals}, r_- : R^{\naturals}, f : (\Mea Y)^X
      &{}\tinf&{}
      \qint_X\parent{\sum_{n \in \naturals} r_n\odot\qmu_n}(\dif x)f(x)
    &{}=
      \sum_{n \in \naturals} r_n\odot \qint_X\qmu_n(\dif x)f(x)
    \\
      \qmu : \Mea X, r_{-} : R^{\naturals}, f_{-} : \carrier\alg^{\naturals \times X}
      &{}\tinf{}&
      \Exp[\alg]_{x \sampled \qmu}\bracks{\sum_{n \in \naturals}r_n\odot f_n(x)}
      &{}=
      \sum_{n \in \naturals}r_n\odot \Exp[\alg]_{x\sampled \qmu}\,[f_n(x)]
    \\\qmu_- : (\Mea X)^{\naturals}, r_- : R^{\naturals}, f : \carrier\alg^X
      &{}\tinf&{}
      \Exp[\alg]_{x \sampled \sum_{n \in \naturals}r_n\odot \qmu_n}[f(x)]
      &{}=
      \sum_{n \in \naturals}r_n\odot \Exp[\alg]_{x \sampled \qmu_n}[f(x)]
  \end{array}
  \]
\end{theorem}

\end{document}